\pdfoutput=1
\documentclass[12pt,a4paper]{article}
\usepackage{ifthen} % for conditional statements
\newboolean{pdflatex}
\setboolean{pdflatex}{true} % False for eps figures 

\newboolean{articletitles}
\setboolean{articletitles}{true} % False removes titles in references

\newboolean{uprightparticles}
\setboolean{uprightparticles}{false} %True for upright particle symbols

\newboolean{inbibliography}
\setboolean{inbibliography}{false} %True once you enter the bibliography

\newboolean{uselhcbcpconvention}
\setboolean{uselhcbcpconvention}{true}

\usepackage{microtype}
\usepackage{lineno}  % for line numbering during review
\usepackage{xspace} % To avoid problems with missing or double spaces after
\usepackage{caption} %these three command get the figure and table captions automatically small

\usepackage{graphicx}  % to include figures (can also use other packages)
\usepackage{color}
\usepackage{colortbl}
\graphicspath{{./figs/}} % Make Latex search fig subdir for figures

\usepackage{amsmath} % Adds a large collection of math symbols
\usepackage{amssymb}
\usepackage{amsfonts}
\usepackage{upgreek} % Adds in support for greek letters in roman typeset

\newcommand*\patchAmsMathEnvironmentForLineno[1]{%
\expandafter\let\csname old#1\expandafter\endcsname\csname #1\endcsname
\expandafter\let\csname oldend#1\expandafter\endcsname\csname
end#1\endcsname
 \renewenvironment{#1}%
   {\linenomath\csname old#1\endcsname}%
   {\csname oldend#1\endcsname\endlinenomath}%
}
\newcommand*\patchBothAmsMathEnvironmentsForLineno[1]{%
  \patchAmsMathEnvironmentForLineno{#1}%
  \patchAmsMathEnvironmentForLineno{#1*}%
}
\AtBeginDocument{%
\patchBothAmsMathEnvironmentsForLineno{equation}%
\patchBothAmsMathEnvironmentsForLineno{align}%
\patchBothAmsMathEnvironmentsForLineno{flalign}%
\patchBothAmsMathEnvironmentsForLineno{alignat}%
\patchBothAmsMathEnvironmentsForLineno{gather}%
\patchBothAmsMathEnvironmentsForLineno{multline}%
\patchBothAmsMathEnvironmentsForLineno{eqnarray}%
}

\usepackage{hyperref}    % Hyperlinks in references
\usepackage[all]{hypcap} % Internal hyperlinks to floats.

\def\lhcb {\mbox{LHCb}\xspace}

\ifthenelse{\boolean{uprightparticles}}%
{

 \def\Ppi         {\ensuremath{\uppi}\xspace}

 \def\Ppsi        {\ensuremath{\uppsi}\xspace}

 \def\PDelta      {\ensuremath{\Delta}\xspace}                 
 \def\PXi      {\ensuremath{\Xi}\xspace}                 
 \def\PLambda      {\ensuremath{\Lambda}\xspace}                 
 \def\PSigma      {\ensuremath{\Sigma}\xspace}                 
 \def\POmega      {\ensuremath{\Omega}\xspace}                 
 \def\PUpsilon      {\ensuremath{\Upsilon}\xspace}                 
 
 \def\PB      {\ensuremath{\mathrm{B}}\xspace}                 
                  
 \def\PD      {\ensuremath{\mathrm{D}}\xspace}

 \def\PJ      {\ensuremath{\mathrm{J}}\xspace}                 
 \def\PK      {\ensuremath{\mathrm{K}}\xspace}

 \def\Pb      {\ensuremath{\mathrm{b}}\xspace}                 
 \def\Pc      {\ensuremath{\mathrm{c}}\xspace}                 
 \def\Pd      {\ensuremath{\mathrm{d}}\xspace}

 \def\Pi      {\ensuremath{\mathrm{i}}\xspace}

 \def\Pp      {\ensuremath{\mathrm{p}}\xspace}

 \def\Ps      {\ensuremath{\mathrm{s}}\xspace}

}
{

 \def\Ppi         {\ensuremath{\pi}\xspace}

 \def\Ppsi        {\ensuremath{\psi}\xspace}                 
                  
 \mathchardef\PDelta="7101
 \mathchardef\PXi="7104
 \mathchardef\PLambda="7103
 \mathchardef\PSigma="7106
 \mathchardef\POmega="710A
 \mathchardef\PUpsilon="7107
                  
 \def\PB      {\ensuremath{B}\xspace}                 
                  
 \def\PD      {\ensuremath{D}\xspace}

 \def\PJ      {\ensuremath{J}\xspace}                 
 \def\PK      {\ensuremath{K}\xspace}

 \def\Pb      {\ensuremath{b}\xspace}                 
 \def\Pc      {\ensuremath{c}\xspace}                 
 \def\Pd      {\ensuremath{d}\xspace}

 \def\Pi      {\ensuremath{i}\xspace}

 \def\Pp      {\ensuremath{p}\xspace}

 \def\Ps      {\ensuremath{s}\xspace}

}

\makeatletter
\ifcase \@ptsize \relax% 10pt
  \newcommand{\miniscule}{\@setfontsize\miniscule{4}{5}}% \tiny: 5/6
\or% 11pt
  \newcommand{\miniscule}{\@setfontsize\miniscule{5}{6}}% \tiny: 6/7
\or% 12pt
  \newcommand{\miniscule}{\@setfontsize\miniscule{5}{6}}% \tiny: 6/7
\fi
\makeatother

\DeclareRobustCommand{\optbar}[1]{\shortstack{{\miniscule (\rule[.5ex]{1.25em}{.18mm})}
  \\ [-.7ex] $#1$}}

\def\g      {{\ensuremath{\Pgamma}}\xspace}

\def\dquark    {{\ensuremath{\Pd}}\xspace}

\def\squark    {{\ensuremath{\Ps}}\xspace}

\def\cquark    {{\ensuremath{\Pc}}\xspace}

\def\bquark    {{\ensuremath{\Pb}}\xspace}

\def\pion   {{\ensuremath{\Ppi}}\xspace}

\def\pip    {{\ensuremath{\pion^+}}\xspace}
\def\pim    {{\ensuremath{\pion^-}}\xspace}

\def\kaon    {{\ensuremath{\PK}}\xspace}
  \def\Kbar    {{\kern 0.2em\overline{\kern -0.2em \PK}{}}\xspace}

\def\KorKbar    {\kern 0.18em\optbar{\kern -0.18em K}{}\xspace}

\def\Kp      {{\ensuremath{\kaon^+}}\xspace}
\def\Km      {{\ensuremath{\kaon^-}}\xspace}

\def\Kstarz  {{\ensuremath{\kaon^{*0}}}\xspace}

  \def\Dbar    {{\kern 0.2em\overline{\kern -0.2em \PD}{}}\xspace}
\def\D       {{\ensuremath{\PD}}\xspace}

\def\DorDbar    {\kern 0.18em\optbar{\kern -0.18em D}{}\xspace}
\def\Dz      {{\ensuremath{\D^0}}\xspace}
\def\Dzb     {{\ensuremath{\Dbar{}^0}}\xspace}

\def\Dm      {{\ensuremath{\D^-}}\xspace}

\def\Dstarp  {{\ensuremath{\D^{*+}}}\xspace}

\def\Ds      {{\ensuremath{\D^+_\squark}}\xspace}
\def\Dsp     {{\ensuremath{\D^+_\squark}}\xspace}
\def\Dsm     {{\ensuremath{\D^-_\squark}}\xspace}

\def\Dssm    {{\ensuremath{\D^{*-}_\squark}}\xspace}

\def\B       {{\ensuremath{\PB}}\xspace}
\def\Bbar    {{\ensuremath{\kern 0.18em\overline{\kern -0.18em \PB}{}}}\xspace}

\def\BorBbar    {\kern 0.18em\optbar{\kern -0.18em B}{}\xspace}
\def\Bz      {{\ensuremath{\B^0}}\xspace}

\def\Bu      {{\ensuremath{\B^+}}\xspace}

\def\Bp      {{\ensuremath{\Bu}}\xspace}

\def\Bd      {{\ensuremath{\B^0}}\xspace}
\def\Bs      {{\ensuremath{\B^0_\squark}}\xspace}
\def\Bsb     {{\ensuremath{\Bbar{}^0_\squark}}\xspace}

\def\jpsi     {{\ensuremath{{\PJ\mskip -3mu/\mskip -2mu\Ppsi\mskip 2mu}}}\xspace}

  \def\Y#1S{\ensuremath{\PUpsilon{(#1S)}}\xspace}% no space before {...}!

\def\proton      {{\ensuremath{\Pp}}\xspace}
\def\antiproton  {{\ensuremath{\overline \proton}}\xspace}

\def\Lz          {{\ensuremath{\PLambda}}\xspace}
\def\Lbar        {{\ensuremath{\kern 0.1em\overline{\kern -0.1em\PLambda}}}\xspace}
\def\LorLbar    {\kern 0.18em\optbar{\kern -0.18em \PLambda}{}\xspace}

\def\Lb      {{\ensuremath{\Lz^0_\bquark}}\xspace}
\def\Lbbar   {{\ensuremath{\Lbar{}^0_\bquark}}\xspace}
\def\Lc      {{\ensuremath{\Lz^+_\cquark}}\xspace}
\def\Lcbar   {{\ensuremath{\Lbar{}^-_\cquark}}\xspace}

\newcommand{\decay}[2]{\ensuremath{#1\!\to #2}\xspace}         % {\Pa}{\Pb \Pc}

\def\to                 {\ensuremath{\rightarrow}\xspace}

\def\CP                {{\ensuremath{C\!P}}\xspace}

\newcommand{\dms}{{\ensuremath{\Delta m_{\squark}}}\xspace}
\newcommand{\dmd}{{\ensuremath{\Delta m_{\dquark}}}\xspace}

\newcommand{\DGs}{{\ensuremath{\Delta\Gamma_{\squark}}}\xspace}

\newcommand{\Gs}{{\ensuremath{\Gamma_{\squark}}}\xspace}

\newcommand{\phis}{{\ensuremath{\phi_{\squark}}}\xspace}
\newcommand{\betas}{{\ensuremath{\beta_{\squark}}}\xspace}

\newcommand{\mistag}{\ensuremath{\omega}\xspace}

\newcommand{\etag}{{\ensuremath{\varepsilon_{\rm tag}}}\xspace}

\newcommand{\effeff}{\ensuremath{\varepsilon_{\rm eff}}\xspace}

\def\AT#1     {\ensuremath{A_{\mathrm{T}}^{#1}}\xspace}           % 2

\def\C#1      {\ensuremath{\mathcal{C}_{#1}}\xspace}                       % 9
\def\Cp#1     {\ensuremath{\mathcal{C}_{#1}^{'}}\xspace}                    % 7
\def\Ceff#1   {\ensuremath{\mathcal{C}_{#1}^{\mathrm{(eff)}}}\xspace}        % 9  
\def\Cpeff#1  {\ensuremath{\mathcal{C}_{#1}^{'\mathrm{(eff)}}}\xspace}       % 7
\def\Ope#1    {\ensuremath{\mathcal{O}_{#1}}\xspace}                       % 2
\def\Opep#1   {\ensuremath{\mathcal{O}_{#1}^{'}}\xspace}                    % 7

\newcommand{\tev}{\ifthenelse{\boolean{inbibliography}}{\ensuremath{~T\kern -0.05em eV}\xspace}{\ensuremath{\mathrm{\,Te\kern -0.1em V}}}\xspace}
\newcommand{\gev}{\ensuremath{\mathrm{\,Ge\kern -0.1em V}}\xspace}
\newcommand{\mev}{\ensuremath{\mathrm{\,Me\kern -0.1em V}}\xspace}
\newcommand{\kev}{\ensuremath{\mathrm{\,ke\kern -0.1em V}}\xspace}
\newcommand{\ev}{\ensuremath{\mathrm{\,e\kern -0.1em V}}\xspace}
\newcommand{\gevc}{\ensuremath{{\mathrm{\,Ge\kern -0.1em V\!/}c}}\xspace}
\newcommand{\mevc}{\ensuremath{{\mathrm{\,Me\kern -0.1em V\!/}c}}\xspace}
\newcommand{\gevcc}{\ensuremath{{\mathrm{\,Ge\kern -0.1em V\!/}c^2}}\xspace}
\newcommand{\gevgevcccc}{\ensuremath{{\mathrm{\,Ge\kern -0.1em V^2\!/}c^4}}\xspace}
\newcommand{\mevcc}{\ensuremath{{\mathrm{\,Me\kern -0.1em V\!/}c^2}}\xspace}

\def\mum  {\ensuremath{{\,\upmu\rm m}}\xspace}

\def\fb   {\ensuremath{\mbox{\,fb}}\xspace}
\def\invfb   {\ensuremath{\mbox{\,fb}^{-1}}\xspace}

\def\ps   {\ensuremath{{\rm \,ps}}\xspace}

\def\invps{\ensuremath{{\rm \,ps^{-1}}}\xspace}

\newcommand{\stat}{\ensuremath{\mathrm{\,(stat)}}\xspace}
\newcommand{\syst}{\ensuremath{\mathrm{\,(syst)}}\xspace}

\def\gsim{{~\raise.15em\hbox{$>$}\kern-.85em
          \lower.35em\hbox{$\sim$}~}\xspace}
\def\lsim{{~\raise.15em\hbox{$<$}\kern-.85em
          \lower.35em\hbox{$\sim$}~}\xspace}

\newcommand{\Real}{\ensuremath{\mathcal{R}e}\xspace}
\newcommand{\Imag}{\ensuremath{\mathcal{I}m}\xspace}

\def\sPlot{\mbox{\em sPlot}}

\def\sqs   {\ensuremath{\protect\sqrt{s}}\xspace}

\def\ptot       {\mbox{$p$}\xspace}
\def\pt         {\mbox{$p_{\rm T}$}\xspace}

\def\rad{\ensuremath{\rm \,rad}\xspace}

\def\evtgen     {\mbox{\textsc{EvtGen}}\xspace}

\def\geant      {\mbox{\textsc{Geant4}}\xspace}

\def\photos     {\mbox{\textsc{Photos}}\xspace}

\def\pythia     {\mbox{\textsc{Pythia}}\xspace}

\def\tell1  {TELL1\xspace}
\def\ukl1   {UKL1\xspace}

\newcommand{\ie}{\mbox{\itshape i.e.}\xspace}

\renewcommand{\Ds}	{\texorpdfstring{\ensuremath{D_{\hspace{-0.0625em}s}}}{Ds}\xspace} % remove charge from LHCb's definition

\newcommand{\BdDK}     {\texorpdfstring{\decay{\Bz}{\Dm \Kp}}{}}
\newcommand{\BdDPi}    {\texorpdfstring{\decay{\Bz}{\Dm \pip}}{}}

\newcommand{\BdDsK}    {\texorpdfstring{\decay{\Bz}{\Dsm \Kp}}{}}
\newcommand{\BdDsPi}   {\texorpdfstring{\decay{\Bz}{\Dsm \pip}}{}}

\newcommand{\BsDsK}    {\texorpdfstring{\decay{\Bs}{\Ds^\mp K^\pm}}{}}
\newcommand{\BsDsPi}   {\texorpdfstring{\decay{\Bs}{\Dsm \pip}}{}}
\newcommand{\BsDsRho}  {\decay{\Bs}{\Dsm \rho^{+}}}

\newcommand{\BsDspKm}  {\texorpdfstring{\decay{\Bs}{\Ds^+ K^-}}{}}
\newcommand{\BsDsmKp}  {\texorpdfstring{\decay{\Bs}{\Ds^- K^+}}{}}
\newcommand{\BsbDspKm}  {\texorpdfstring{\decay{\Bsb}{\Ds^+ K^-}}{}}

\newcommand{\BsDsstPi} {\decay{\Bs}{\Dssm\pip}}

\newcommand{\DsKKPi}   {\decay{\Dsm}{\Km\Kp\pim}}
\newcommand{\DsKPiPi}  {\decay{\Dsm}{\Km\pip\pim}}

\newcommand{\DsPiPiPi} {\decay{\Dsm}{\pim\pip\pim}}
\newcommand{\LbDsOrDsstp}{\decay{\Lb}{\D^{(*)-}_\squark p}}
\newcommand{\LbDsP}    {\texorpdfstring{\decay{\Lb}{\Dsm p}}{}}
\newcommand{\LbDsstP}  {\decay{\Lb}{\Dssm p}}
\newcommand{\LbLcK}    {\decay{\Lbbar}{\Lcbar \Kp}}
\newcommand{\LbLcPi}   {\decay{\Lbbar}{\Lcbar \pip}}

\newcommand{\DsK}      {\ensuremath{\Ds\hspace{-0.0625em} K}\xspace}

\newcommand{\DsmKp}    {\ensuremath{\Dsm\Kp}\xspace}
\newcommand{\DspKm}    {\ensuremath{\Dsp\Km}\xspace}

\renewcommand{\g}	{\ensuremath{\gamma}\xspace}
\renewcommand{\betas}	{\ensuremath{\beta_s}\xspace}
\newcommand{\deltams}	{\ensuremath{\Delta m_s}\xspace}
\renewcommand{\dms}	{\deltams}
\newcommand{\gs}	{\ensuremath{\Gamma_s}\xspace}
\newcommand{\dgs}	{\ensuremath{\Delta\Gamma_s}\xspace}

\newcommand{\f}		{\ensuremath{f}\xspace} % the final state, e.g. DsK or simply f
\renewcommand{\fb}	{\ensuremath{\overline{\f}}\xspace} % redefining LHCb's fb for femtobarn
\newcommand{\Af}	{\ensuremath{A_{\f}}\xspace} % the decay amplitude to final state f

\newcommand{\Abf}	{\ensuremath{\overline{A}_{\f}}\xspace}

\newcommand{\weak}	{\ensuremath{\gamma - 2\betas}\xspace}
\newcommand{\strong}	{\ensuremath{\delta}\xspace}
\newcommand{\lf}	{\ensuremath{\lambda_{\f}}\xspace}

\newcommand{\lfb}	{\ensuremath{\lambda_{\fb}}\xspace}
\newcommand{\rdsk}	{\ensuremath{r_{\DsK}}\xspace}
\newcommand{\omcl}	{\ensuremath{1-{\rm CL}}\xspace}

\ifthenelse{\boolean{uselhcbcpconvention}}%
{%
\newcommand{\Cbpar}	{\ensuremath{C_{\fb}}\xspace}
\newcommand{\Cpar}	{\ensuremath{C_{\f}}\xspace}
\newcommand{\Sbpar}	{\ensuremath{S_{\fb}}\xspace}
\newcommand{\Spar}	{\ensuremath{S_{\f}}\xspace}
\newcommand{\Dbpar}	{\ensuremath{{A_{\fb}^{\Delta\Gamma}}}\xspace}
\newcommand{\Dpar}	{\ensuremath{{A_{\f}^{\Delta\Gamma}}}\xspace}

}{%
\newcommand{\Cbpar}	{\ensuremath{C_{\fb}}\xspace}
\newcommand{\Cpar}	{\ensuremath{C_{\f}}\xspace}
\newcommand{\Sbpar}	{\ensuremath{S_{\fb}}\xspace}
\newcommand{\Spar}	{\ensuremath{S_{\f}}\xspace}
\newcommand{\Dbpar}	{\ensuremath{D_{\fb}}\xspace}
\newcommand{\Dpar}	{\ensuremath{D_{\f}}\xspace}

}%

\newcommand{\pidk}{\ensuremath{L(K/\pi)}\xspace}

\newcommand{\sfit}	{\mbox{\it sFit}\xspace}
\newcommand{\cfit}	{\mbox{\it cFit}\xspace}
\renewcommand{\sPlot}	{\mbox{\it sPlot}\xspace}
\newcommand{\sWeights}	{\mbox{\it sWeights}\xspace}
\newcommand{\sWeighted}	{\mbox{\it sWeighted}\xspace}

\usepackage{cite} % Allows for ranges in citations
\usepackage{mciteplus}

\begin{document}

%%%%%%%%%%%%%%%%%%%%%%%%%
%%%%% Title     %%%%%%%%%
%%%%%%%%%%%%%%%%%%%%%%%%%
\renewcommand{\thefootnote}{\fnsymbol{footnote}}
\setcounter{footnote}{1}

% %%%%%%% CHOOSE TITLE PAGE--------
%\onecolumn
% $Id: title-LHCb-PAPER.tex 52263 2014-04-15 19:53:37Z roldeman $
% ===============================================================================
% Purpose: LHCb-PAPER journal paper title page template
% Author: 
% Created on: 2010-09-25
% ===============================================================================

%%%%%%%%%%%%%%%%%%%%%%%%%
%%%%%  TITLE PAGE  %%%%%%
%%%%%%%%%%%%%%%%%%%%%%%%%
\begin{titlepage}
\pagenumbering{roman}

% Header ---------------------------------------------------
\vspace*{-1.5cm}
\centerline{\large EUROPEAN ORGANISATION FOR NUCLEAR RESEARCH (CERN)}
\vspace*{1.5cm}
\hspace*{-0.5cm}
\begin{tabular*}{\linewidth}{lc@{\extracolsep{\fill}}r}
\ifthenelse{\boolean{pdflatex}}% Logo format choice
{\vspace*{-1.2cm}\mbox{\!\!\!\includegraphics[width=.14\textwidth]{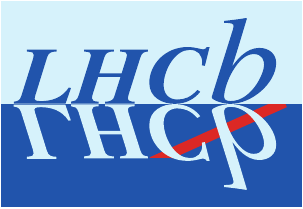}} & &}%
{\vspace*{-1.2cm}\mbox{\!\!\!\includegraphics[width=.12\textwidth]{lhcb-logo.eps}} & &}%
\\
 & & CERN-PH-EP-2014-168 \\  % ID 
 & & LHCb-PAPER-2014-038 \\  % ID 
 & & July 23, 2014 \\ % Date - Can also hardwire e.g.: 23 March 2010
 & & \\
% not in paper \hline
\end{tabular*}

\vspace*{2.0cm}

% Title --------------------------------------------------
{\bf\boldmath\huge
\begin{center}
	Measurement of \CP asymmetry in \BsDsK decays
\end{center}
}

\vspace*{1.0cm}

% Authors -------------------------------------------------
\begin{center}
The LHCb collaboration\footnote{Authors are listed at the end of this article.}
\end{center}

\vspace{\fill}

% Abstract -----------------------------------------------
\begin{abstract}
  \noindent
  We report on measurements of the time-dependent \CP violating observables
in \BsDsK decays using a dataset corresponding to 1.0\invfb of $pp$ collisions recorded
with the \lhcb detector.
We find the \CP violating observables  
$\Cpar  =            0.53 \pm 0.25 \pm 0.04$, 
$\Dpar  =            0.37 \pm 0.42 \pm 0.20$,
$\Dbpar =            0.20 \pm 0.41 \pm 0.20$,
$\Spar  =           -1.09 \pm 0.33 \pm 0.08$,
$\Sbpar =           -0.36 \pm 0.34 \pm 0.08$, 
where the uncertainties are statistical and 
systematic, respectively. Using these observables together with a recent
measurement of the $\Bs$ mixing phase $-2\beta_s$ leads to the first
extraction of the CKM angle $\gamma$ from \BsDsK decays, finding $\gamma =
(115_{-43}^{+28})^\circ$~modulo~$180^\circ$ at 68\% CL, where the error
contains both statistical and systematic uncertainties.
\end{abstract}
\vspace*{1.0cm}

\begin{center}
    Published in \href{http://dx.doi.org/10.1007/JHEP11(2014)060}{JHEP 11 (2014) 060}
\end{center}

\vspace{\fill}

{\footnotesize 
\centerline{\copyright~CERN on behalf of the \lhcb collaboration, license \href{http://creativecommons.org/licenses/by/4.0/}{CC-BY-4.0}.}}
\vspace*{2mm}

\end{titlepage}

%%%%%%%%%%%%%%%%%%%%%%%%%%%%%%%%
%%%%%  EOD OF TITLE PAGE  %%%%%%
%%%%%%%%%%%%%%%%%%%%%%%%%%%%%%%%

%  empty page follows the title page ----
\newpage
\setcounter{page}{2}
\mbox{~}

\cleardoublepage

%\twocolumn
% %%%%%%%%%%%%% ---------

\renewcommand{\thefootnote}{\arabic{footnote}}
\setcounter{footnote}{0}

%%%%%%%%%%%%%%%%%%%%%%%%%%%%%%%%
%%%%%  Table of Content   %%%%%%
%%%%%%%%%%%%%%%%%%%%%%%%%%%%%%%%
%%%% Uncomment next 2 lines if desired
%\tableofcontents
%\cleardoublepage

%%%%%%%%%%%%%%%%%%%%%%%%%
%%%%% Main text %%%%%%%%%
%%%%%%%%%%%%%%%%%%%%%%%%%

\pagestyle{plain} % restore page numbers for the main text
\setcounter{page}{1}
\pagenumbering{arabic}

%% Uncomment during review phase. 
%% Comment before a final submission.
%\linenumbers

% You can include short sections directly in the main tex file.
% However, for larger papers it is desirable to split the text into
% several semiautonomous files, which can be revised independently.
% This is especially useful when developing a document in
% collaboration with several people, since then different parts can be
% edited independently.  This type of file organization is shown here.
% 

\section{Introduction}
% --------------------
\label{sec:intro}
Time-dependent analyses of
tree-level $\B^0_{(s)}\to \D^\mp_{(s)} {\pi^\pm,K^\pm}$ decays\footnote{Inclusion of charge
conjugate modes is implied except where explicitly stated.}
are sensitive to the angle $\g\equiv\arg(-V_{ud}V_{ub}^{*}/V_{cd}V_{cb}^{*})$ of the unitarity triangle of the Cabibbo-Kobayashi-Maskawa (CKM)
matrix \cite{CKM1,CKM2} through \CP violation in the interference of mixing and decay amplitudes~\cite{Dunietz:1987bv,Aleksan:1991nh,Fleischer:2003yb}.
The determination of \g from such tree-level decays is important because it is not sensitive
to potential effects from most models of physics beyond the Standard Model (BSM). The value of \g hence provides 
a reference against which other BSM-sensitive measurements can be compared.

Due to the interference between mixing and decay amplitudes, the physical \CP violating observables in these decays
are functions of  a combination of \g and the relevant mixing phase,
namely $\gamma+2\beta$ ($\beta \equiv \arg(-V_{cd}V_{cb}^{*}/V_{td}V_{tb}^{*})$)
in the \Bd and \weak
($\betas \equiv \arg(-V_{ts}V_{tb}^{*}/V_{cs}V_{cb}^{*})$)
in the \Bs system. A measurement of these physical
observables can therefore be interpreted in terms of \g or $\beta_{(s)}$
by using an independent measurement of the other parameter as input.

Such measurements have been performed by both the BaBar~\cite{Aubert:2005yf,Aubert:2006tw}
and the Belle~\cite{PhysRevD.73.092003,Bahinipati:2011yq} collaborations
using $\Bd\to D^{(*)\mp}\pi^\pm$ decays. In these decays, however, the 
ratios $r_{D^{(*)}\pi} = |A(\Bd \to D^{(*)-}\pi^+)/A(\Bd \to D^{(*)+}\pi^-)|$
between the interfering $b \to u$ and $b \to c$ amplitudes
are small, $r_{D^{(*)}\pi} \approx 0.02$, limiting the
sensitivity on \g~\cite{Baak:2007gp}.

\begin{figure}[h]
  \centering % figures a tad smaller to keed them on the same page
  $\,$ \hfill \includegraphics[width=.42\textwidth]{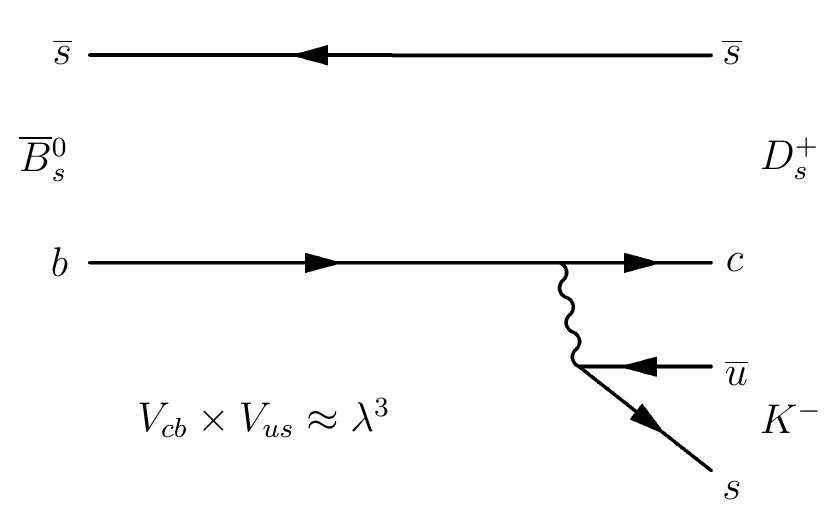} \hfill
  \includegraphics[width=.42\textwidth]{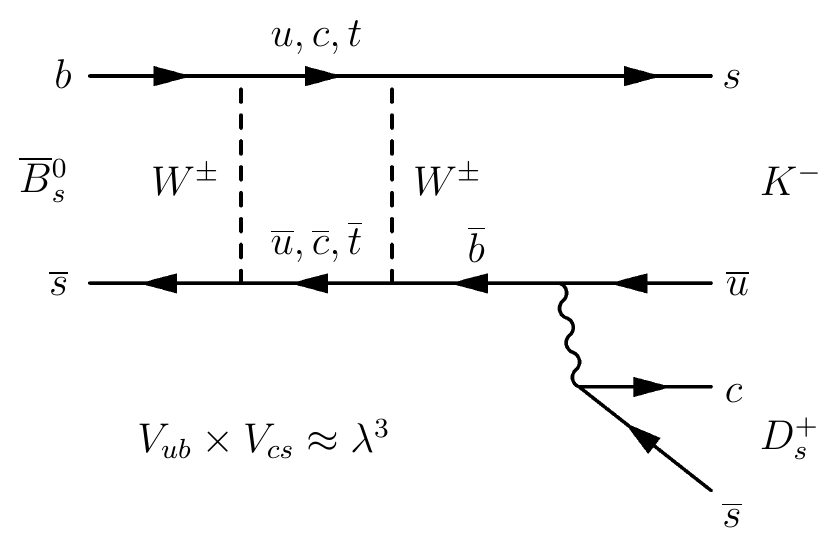} \hfill $\,$ \\
\vspace{-2mm}
  \caption{Feynman diagrams for \BsbDspKm without (left) and with (right) \Bs mixing.}
  \label{fig:feynmandiags}
\end{figure}

The leading order Feynman diagrams contributing to the interference of decay
and mixing in \BsDsK are shown in Fig.~\ref{fig:feynmandiags}. In contrast to $\Bd\to D^{(*)\mp}\pi^\pm$ decays, here both the
\BsDsmKp ($b\to cs\bar{u}$) and \BsDspKm ($b\to u\bar{c}s$)
amplitudes are of the same order in the sine of the Cabibbo angle $\lambda = 0.2252 \pm 0.0007$~\cite{Wolfenstein:1983yz,PDG},
$\mathcal{O}(\lambda^3)$, and the amplitude ratio of the interfering diagrams is approximately $|V_{ub}V_{cs}/V_{cb}V_{us}| \approx 0.4$.
Moreover, the decay width difference in the \Bs system,
\dgs, is nonzero~\cite{LHCb-PAPER-2013-002},
which allows a determination of \weak
from the sinusoidal and hyperbolic terms
in the decay time evolution, up to a two-fold ambiguity.

This paper presents the first measurements of the
\CP violating observables in \BsDsK decays using a dataset corresponding to 1.0 \invfb of  
$pp$ collisions recorded with the \lhcb detector at $\sqs = 7\tev$, and the first determination 
of \weak in these decays.
\vspace{-2mm}

\subsection{Decay rate equations and \CP violation observables}
\label{sec:equations}
The time-dependent decay rates of the initially produced flavour eigenstates
$|\Bs(t=0)\rangle$ and $|\Bsb(t=0)\rangle$ are given
by
\begin{align}
\frac{{\rm d}\Gamma_{\Bs\to\f}(t)}{{\rm d}t} &= \frac{1}{2} |\Af|^2 (1+|\lf|^2) e^{-\gs t} \left[
         \cosh\left(\frac{\dgs t}{2}\right) 
  + \Dpar\sinh\left(\frac{\dgs t}{2}\right) \right. \nonumber\\
& + \Cpar\cos\left(\dms t\right) 
  - \Spar\sin\left(\dms t\right)
\Big],
\label{eq:decay_rates_1}\\
\frac{{\rm d}\Gamma_{\Bsb\to\f}(t)}{{\rm d}t} &= \frac{1}{2} |\Af|^2 \left|\frac{p}{q}\right|^2 (1+|\lf|^2) e^{-\gs t} \left[
         \cosh\left(\frac{\dgs t}{2}\right) 
  + \Dpar\sinh\left(\frac{\dgs t}{2}\right) \right. \nonumber\\
& - \Cpar\cos\left(\dms t\right) 
  + \Spar\sin\left(\dms t\right)
\Big],
\label{eq:decay_rates_2}
\end{align}
where \mbox{$\lf \equiv (q/p)(\Abf/\Af)$} and \Af (\Abf)
is the decay amplitude of a
\Bs to decay to a final state \f (\fb).
\gs is the average \Bs decay width, and \dgs is the positive~\cite{LHCb-PAPER-2011-028}
decay-width difference between the heavy and light mass eigenstates in the \Bs system.
The complex coefficients $p$ and $q$ relate the \Bs meson mass
eigenstates, $|B_{L,H}\rangle$, to the flavour eigenstates,
$|\Bs\rangle$ and $|\Bsb\rangle$
\begin{align}
|B_L\rangle &= p|\Bs\rangle+q|\Bsb\rangle\,, \\
|B_H\rangle &= p|\Bs\rangle-q|\Bsb\rangle\,,
\label{eq:mixing}
\end{align}
with $|p|^2+|q|^2=1$.
Similar equations can be written for the \CP-conjugate decays replacing
\Cpar by \Cbpar, \Spar by \Sbpar, and
\Dpar by \Dbpar. In our convention \f is the $\Dsm \Kp$ final state and \fb is $\Dsp \Km$.
The \CP asymmetry observables \Cpar, \Spar, \Dpar, \Cbpar,
\Sbpar and \Dbpar are given by
\begin{align}
\Cpar  =  \frac{ 1 - |\lf|^2 }{ 1 + |\lf|^2 } & = -\Cbpar = - \frac{1-|\lfb|^2}{1+|\lfb|^2}\,, \nonumber\\
\Spar  = \frac{ 2 \Imag(\lf) }  { 1 + |\lf|^2 }        \,&,\quad
\Dpar  = \frac{ -2 \Real(\lf) }  { 1 + |\lf|^2 }        \,,  \nonumber\\
\Sbpar = \frac{ 2 \Imag(\lfb) }{ 1 + |\lfb|^2 }      \,&, \quad
\Dbpar = \frac{ -2 \Real(\lfb) }{ 1 + |\lfb|^2 }      \,.
\label{eq:asymm_obs}
\end{align}
The equality $\Cpar = -\Cbpar$ results from $|q/p| = 1$ and
$|\lf| = |\frac{1}{\lfb}|$, \ie the assumption of no \CP violation in either the decay or mixing amplitudes.
The \CP observables are related to the
magnitude of the amplitude ratio $\rdsk \equiv |\lambda_{\DsK}| = |A(\Bsb \to \Dsm \Kp)/A(\Bs \to \Dsm \Kp)|$,
the strong phase difference \strong, and the weak phase difference \weak by the following equations:
\begin{align}
\Cpar  	= &\frac{1-\rdsk^2}{1+\rdsk^2}                   \,,   \nonumber\\
\Dpar 	= \frac{-2 \rdsk \cos(\strong-(\weak))}{1+\rdsk^2}\,&, \quad
\Dbpar  = \frac{-2 \rdsk \cos(\strong+(\weak))}{1+\rdsk^2}\,,  \nonumber\\
\Spar 	= \frac{2 \rdsk \sin(\strong-(\weak))}{1+\rdsk^2}\,&, \quad
\Sbpar	= \frac{-2 \rdsk \sin(\strong+(\weak))}{1+\rdsk^2}\,.
\label{eq:truth}
\end{align}

\subsection{Analysis strategy}
\label{sec:strategy}

To measure the \CP violating observables defined in
Sec.~\ref{sec:equations}, it is necessary to perform a fit to the decay-time
distribution of the selected \BsDsK candidates. The kinematically similar mode
\BsDsPi is used as control channel which helps in the determination of the
time-dependent efficiency and flavour tagging performance.  Before a fit to the
decay time can be performed, it is necessary to distinguish the signal and
background candidates in the selected sample. This analysis uses three
variables to maximise sensitivity when discriminating between signal and
background: the \Bs mass; the \Dsm mass; and the log-likelihood difference
\pidk between the pion and kaon hypotheses for the companion particle.

In Sec.~\ref{sec:sigbacklineshapes}, the signal and background shapes needed for the analysis are obtained in each
of the variables. Section~\ref{sec:mdfit} describes how a simultaneous extended maximum likelihood fit (in the following referred
to as multivariate fit) 
to these three variables is used to determine the
yields of signal and background components in the samples of \BsDsPi and \BsDsK candidates. Section~\ref{sec:tagging} describes how to obtain the flavour at production
of the \BsDsK candidates using a combination of flavour-tagging algorithms, whose performance
is calibrated with data using flavour-specific control modes. The decay-time resolution
and acceptance are determined using a mixture of data control modes and simulated signal events, described in Sec.~\ref{sec:timeresandacc}.

Finally, Sec.~\ref{sec:timefit} describes two approaches to fit the decay-time distribution of the \BsDsK candidates which extract the
\CP violating observables. The first fit, henceforth referred
to as the \sfit, uses the results of the multivariate fit to obtain 
the so-called \sWeights~\cite{Pivk:2004ty} which
allow the background components to be statistically subtracted~\cite{2009arXiv0905.0724X}.
The \sfit to the decay-time distribution
is therefore performed using only the probability density function (PDF) of the signal component. The second fit, henceforth referred to
as the \cfit, uses the various shapes and yields of the multivariate fit result for the different signal and background components.
The \cfit subsequently performs a six-dimensional maximum likelihood fit to these variables, the decay-time distribution and uncertainty,
and the probability that the initial \Bs flavour is correctly determined,
in which all contributing signal and background components are described with their appropriate PDFs.
In Sec.~\ref{sec:interpretation}, we extract the CKM angle \g using the result of one of the two approaches.

\section{Detector and software}
\label{sec:Detector}
The \lhcb detector~\cite{Alves:2008zz} is a single-arm forward
spectrometer covering the \mbox{pseudorapidity} range $2<\eta <5$,
designed for the study of particles containing \bquark or \cquark
quarks. The detector includes a high-precision tracking system
consisting of a silicon-strip vertex detector surrounding the $pp$
interaction region~\cite{LHCb-DP-2014-001}, a large-area silicon-strip detector located
upstream of a dipole magnet with a bending power of about
$4{\rm\,Tm}$, and three stations of silicon-strip detectors and straw
drift tubes~\cite{LHCb-DP-2013-003} placed downstream of the magnet.
The tracking system provides a measurement of momentum, \ptot,  with
a relative uncertainty that varies from 0.4\% at low momentum to 0.6\% at 100\gevc.
The minimum distance of a track to a primary $pp$ collision vertex, the impact parameter, is measured with a resolution of $(15+29/\pt)\mum$,
where \pt is the component of \ptot transverse to the beam, in \gevc.
Different types of charged hadrons are distinguished using information
from two ring-imaging Cherenkov detectors~\cite{LHCb-DP-2012-003}.
The magnet polarity is reversed regularly to control systematic effects.

The trigger~\cite{LHCb-DP-2012-004} consists of a
hardware stage, based on information from the calorimeter and muon
systems, followed by a software stage, which applies a full event
reconstruction. The software trigger requires a two-, three- or four-track
secondary vertex with a large sum of the transverse momentum of
the charged particles and a significant displacement from the primary $pp$
interaction vertices~(PVs). A multivariate algorithm~\cite{BBDT} is used for
the identification of secondary vertices consistent with the decay
of a \bquark hadron.

In the simulation, $pp$ collisions are generated using
\pythia~\cite{Sjostrand:2006za} 
with a specific \lhcb configuration~\cite{LHCb-PROC-2010-056}.  Decays of hadrons
are described by \evtgen~\cite{Lange:2001uf}, in which final state
radiation is generated using \photos~\cite{Golonka:2005pn}. The
interaction of the generated particles with the detector and its
response are implemented using the \geant
toolkit~\cite{Allison:2006ve, Agostinelli:2002hh} as described in
Ref.~\cite{LHCb-PROC-2011-006}.

\section{Event selection}
\label{sec:selection}

The event selection begins by building 
\DsKKPi, \DsKPiPi, and \DsPiPiPi candidates from reconstructed charged particles. These \Dsm candidates
are subsequently combined with a fourth particle, referred to as the
``companion'', to form \BsDsK and \BsDsPi candidates.
The flavour-specific Cabibbo-favoured decay mode \BsDsPi is used as a control
channel in the analysis, and is selected identically
to \BsDsK except for the PID criteria on the companion particle.
The decay-time and \Bs mass resolutions are improved by performing a kinematic fit~\cite{Hulsbergen2005566} in which the
\Bs candidate is constrained to originate from its associated proton-proton interaction, \ie\xspace the one with the smallest IP with respect to the \Bs candidate,
and the \Bs mass is computed with a constraint on the \Dsm mass.

The \BsDsPi mode is used for the optimisation of the selection and
for studying and constraining physics backgrounds to the \BsDsK decay.
The \BsDsK and \BsDsPi candidates are required to be matched to the
secondary vertex candidates found in the software trigger.
Subsequently, a preselection is applied to the \BsDsK and \BsDsPi candidates
using a similar multivariate displaced vertex algorithm to the trigger selection,
but with offline-quality reconstruction.

A selection using the gradient boosted decision tree (BDTG)~\cite{Breiman}
implementation in the {\sc Tmva} software package~\cite{TMVA} further suppresses combinatorial backgrounds.
The BDTG is trained on data using the 
\BsDsPi, \DsKKPi decay sample, which is purified with
respect to the previous preselection exploiting PID
information from the Cherenkov detectors.
Since all channels in this analysis are kinematically similar, and since no PID information is used as input to the BDTG,
the resulting BDTG performs equally well on the other \Dsm decay modes.
The optimal working point is chosen to maximise the expected sensitivity 
to the \CP violating observables in \BsDsK decays. 
In addition, the \Bs and \Dsm candidates
are required to be within $m(\Bs) \in [5300,5800]\mevcc$ and $m(\Dsm) \in [1930,2015]\mevcc$, respectively.

Finally, the different final states are distinguished by using PID information. 
This selection also
strongly suppresses cross-feed and peaking backgrounds from other misidentified
decays of $b$-hadrons to $c$-hadrons. We will refer to such backgrounds as ``fully reconstructed''
if no particles are missed in the reconstruction, and ``partially reconstructed'' otherwise.
The decay modes \BdDPi, \BdDsPi, \LbLcPi, \BsDsK, and \BsDsstPi are backgrounds to \BsDsPi,
while \BsDsPi, \BsDsstPi, \BsDsRho, \BdDsK, \BdDK, \BdDPi, \LbLcK, \LbLcPi, and \LbDsOrDsstp
are backgrounds to \BsDsK. This part of the selection is necessarily different for
each \Dsm decay mode, as described below. 
\begin{itemize}
\item For \DsPiPiPi none of the possible misidentified backgrounds fall inside the \Dsm mass window. Loose
PID requirements are nevertheless used to identify
the \Dsm decay products as pions in order to suppress combinatorial background.
\item For \DsKPiPi, the relevant peaking backgrounds are $\Lcbar\to\antiproton\pip\pim$ in which the antiproton
is misidentified, and $\Dm \to \Kp\pim\pim$ in which both the kaon and a pion are misidentified. As this is
the smallest branching fraction \Dsm decay mode used, and hence that most affected
by background, all \Dsm decay products are required to pass tight PID requirements.
\item The \DsKKPi mode is split into three submodes.
We distinguish between the resonant $\Dsm\to\phi\pim$ and $\Dsm\to \Kstarz K^-$ decays,
and the remaining decays.
Candidates in which the $K^+K^-$ pair falls within 
20\mevcc of the $\phi$ mass are identified as a $\Dsm\to\phi\pim$ decay. This requirement suppresses most of
the cross-feed and combinatorial background, and only loose PID
requirements are needed. 
Candidates within a 50\mevcc window around the \Kstarz mass
are identified as a $\Dsm\to \Kstarz K^-$ decay; it is kinematically impossible
for a candidate to satisfy both this and the $\phi$ requirement. In this case there 
is non-negligible background from misidentified
$\Dm\to K^+\pim\pim$ and $\Lcbar\to\antiproton\pim\Kp$ decays which are suppressed through
tight PID requirements on the \Dsm kaon with the same charge as the \Dsm pion.
The remaining candidates, referred to as nonresonant decays, are subject to
tight PID requirements on all decay products to suppress cross-feed
backgrounds.
\end{itemize}
\begin{figure}[b]
  \centering
  \includegraphics[width=.49\textwidth]{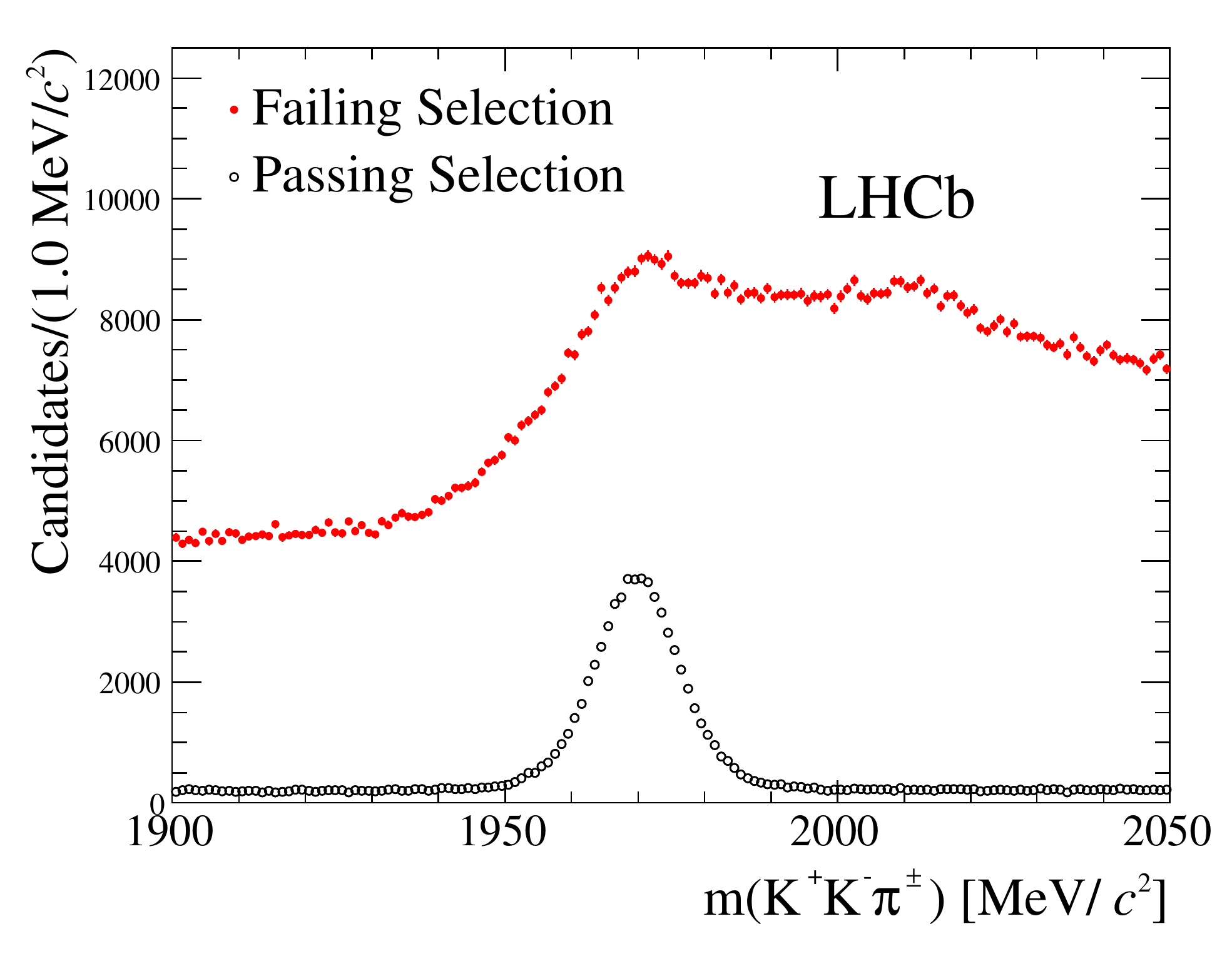}
  \includegraphics[width=.49\textwidth]{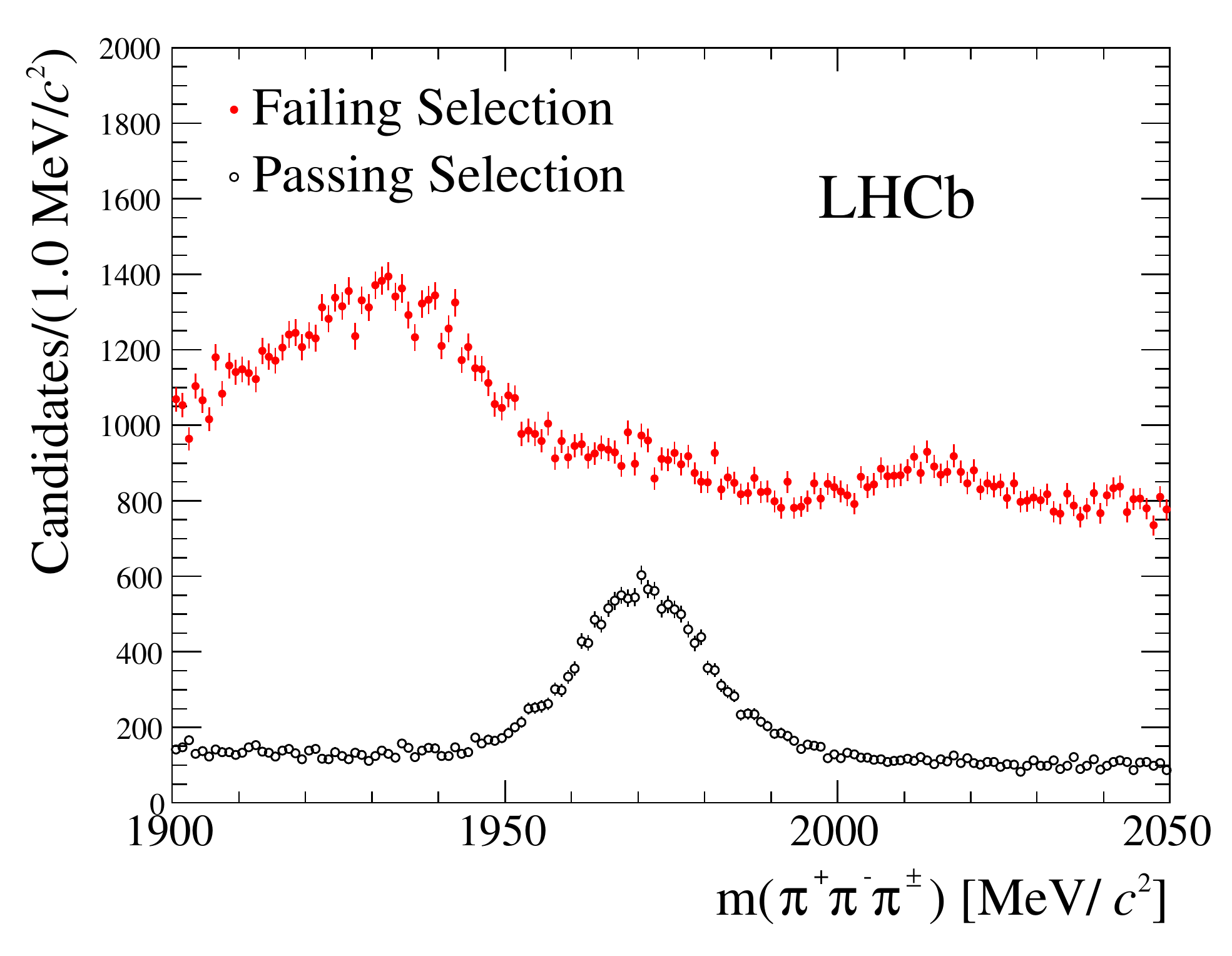} \\
  \includegraphics[width=.49\textwidth]{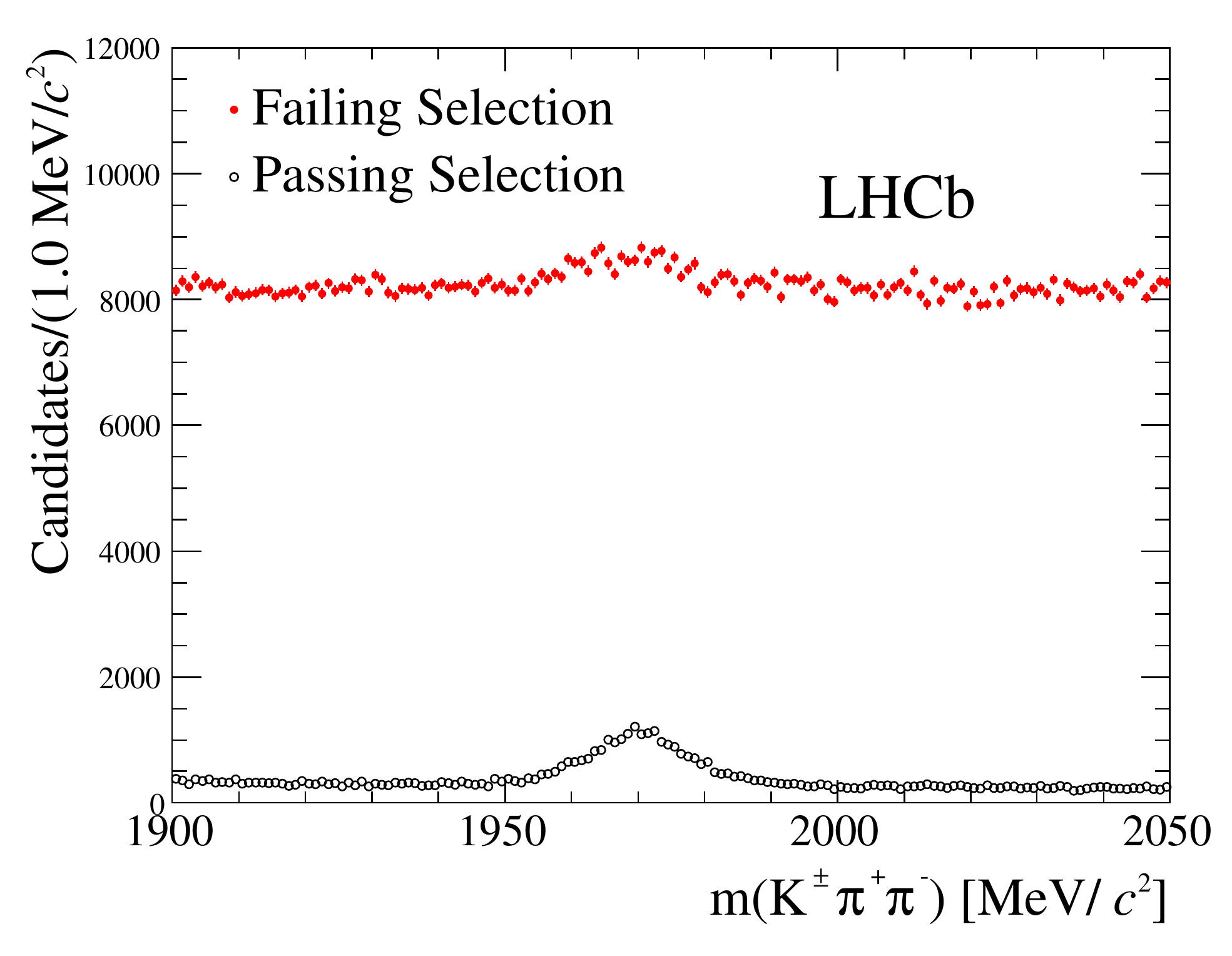}
  \caption{Mass distributions for \Dsm candidates passing (black, open circles) and failing (red, crosses) the PID selection criteria.
In reading order: \DsKKPi, \DsPiPiPi, and \DsKPiPi. }
  \label{fig:massdists-Ds-PrePost}
\end{figure}
\newpage
Figure~\ref{fig:massdists-Ds-PrePost} shows the relevant mass distributions for candidates passing and failing this PID
selection. Finally a loose PID requirement is made on the companion track.
After all selection requirements, fewer than 2$\%$ of retained events contain more than one signal candidate. 
All candidates are used in the subsequent analysis.

\section{Signal and background shapes}
\label{sec:sigbacklineshapes}

The signal and background shapes are obtained using a mixture of data-driven approaches and simulation.
The simulated events need to be corrected for kinematic differences between simulation and data, as well
as for the kinematics-dependent efficiency of the PID selection requirements. In order to obtain
kinematic distributions in data for this weighting, we use the
decay mode \BdDPi, which can be selected with very high purity without the use of any PID requirements
and is kinematically very similar to the \Bs signals.
The PID efficiencies are measured as a function of particle momentum and event occupancy using
prompt $\Dstarp\to\Dz(\Km\pip)\pip$ decays which provide pure samples of pions and kaons~\cite{LHCb-PROC-2011-008}, henceforth called $D^{*+}$ calibration sample.

\subsection{$\mathbf{\Bs}$ candidate mass shapes}

In order to model radiative and reconstruction effects,
the signal shape in the \Bs mass is the sum of two Crystal Ball~\cite{Skwarnicki:1986xj} functions
with common mean and oppositely oriented tails.
The signal shapes are determined separately for \BsDsK and \BsDsPi from simulated candidates.
The shapes are subsequently fixed in the
multivariate fit except for the common mean of the Crystal Ball functions which floats for both the \BsDsPi and \BsDsK channel.

The functional form of the combinatorial background is taken from the upper \Bs sideband,
with its parameters left free to vary in the subsequent multivariate fit.
Each \Dsm mode is considered independently and parameterised by either an exponential function
or by a combination of an exponential and a constant function.

The shapes of the fully or partially reconstructed backgrounds are fixed from
simulated events using a non-parametric kernel estimation
method~(KEYS,~\cite{Cranmer:2000du}). Exceptions to this are the \BdDPi
background in the \BsDsPi fit and the \BsDsPi background in the \BsDsK fit,
which are obtained from data.  The latter two backgrounds are reconstructed
with the ``wrong'' mass hypothesis but without PID requirements, which would
suppress them. The resulting shapes are then weighted to account for the effect
of the momentum-dependent efficiency of the PID requirements from the $D^{*+}$
calibration samples, and KEYS templates are extracted for use in the
multivariate fit.

\subsection{$\mathbf{\Dsm}$ candidate mass shapes}

The signal shape in the \Dsm mass is again a sum of two Crystal Ball functions
with common mean and oppositely oriented tails. 
The signal shapes are extracted separately for each \Dsm decay mode from simulated events that have the full 
selection chain applied to them. The shapes are subsequently fixed in the
multivariate fit except for the common mean of the Crystal Ball functions, which floats independently for each \Dsm decay mode.

The combinatorial background consists of both random combinations of tracks which do not peak in the \Dsm mass, and, in
some \Dsm decay modes, backgrounds that contain a true \Dsm, and a random companion track. It is parameterised separately
for each \Dsm decay mode either by an exponential function or by a combination of an exponential function and the signal \Dsm shape.

The fully and partially reconstructed backgrounds which contain a
correctly reconstructed \Dsm candidate (\BsDsK and \BdDsPi as backgrounds in the
\BsDsPi fit; \BdDsK and \BsDsPi as backgrounds in the \BsDsK fit)
are assumed to have the same mass distribution as the signal. For other backgrounds,
the shapes are KEYS templates taken from simulated events, as in the \Bs mass.

\subsection{Companion $\mathbf{\pidk}$ shapes}

We obtain the PDFs describing the \pidk
distributions of pions and kaons from
dedicated $D^{*+}$ calibration samples. We obtain the PDF describing the protons
using a calibration sample of $\Lc\to p K^- \pi^+$ decays. These samples are weighted to match
the signal kinematic and event occupancy distributions in the same way as the simulated events. 
The weighting is done separately for each signal and
background component, as well as for each magnet polarity.
The shapes for each magnet polarity are subsequently combined according
to the integrated luminosity in each sample.

The signal companion \pidk shape is obtained separately for each \Dsm decay mode to account
for small kinematic differences between them. The combinatorial background companion \pidk shape is taken
to consist of a mixture of pions, protons, and kaons, and its normalisation
is left floating in the multivariate fit. The companion \pidk shape for fully or
partially reconstructed backgrounds is obtained by weighting the PID calibration samples
to match the event distributions of simulated events, for each background type.

\section{Multivariate fit to $\mathbf{\BsDsK}$ and $\mathbf{\BsDsPi}$}
\label{sec:mdfit}

\begin{figure}[htb]
  \centering
  \includegraphics[width=.49\textwidth]{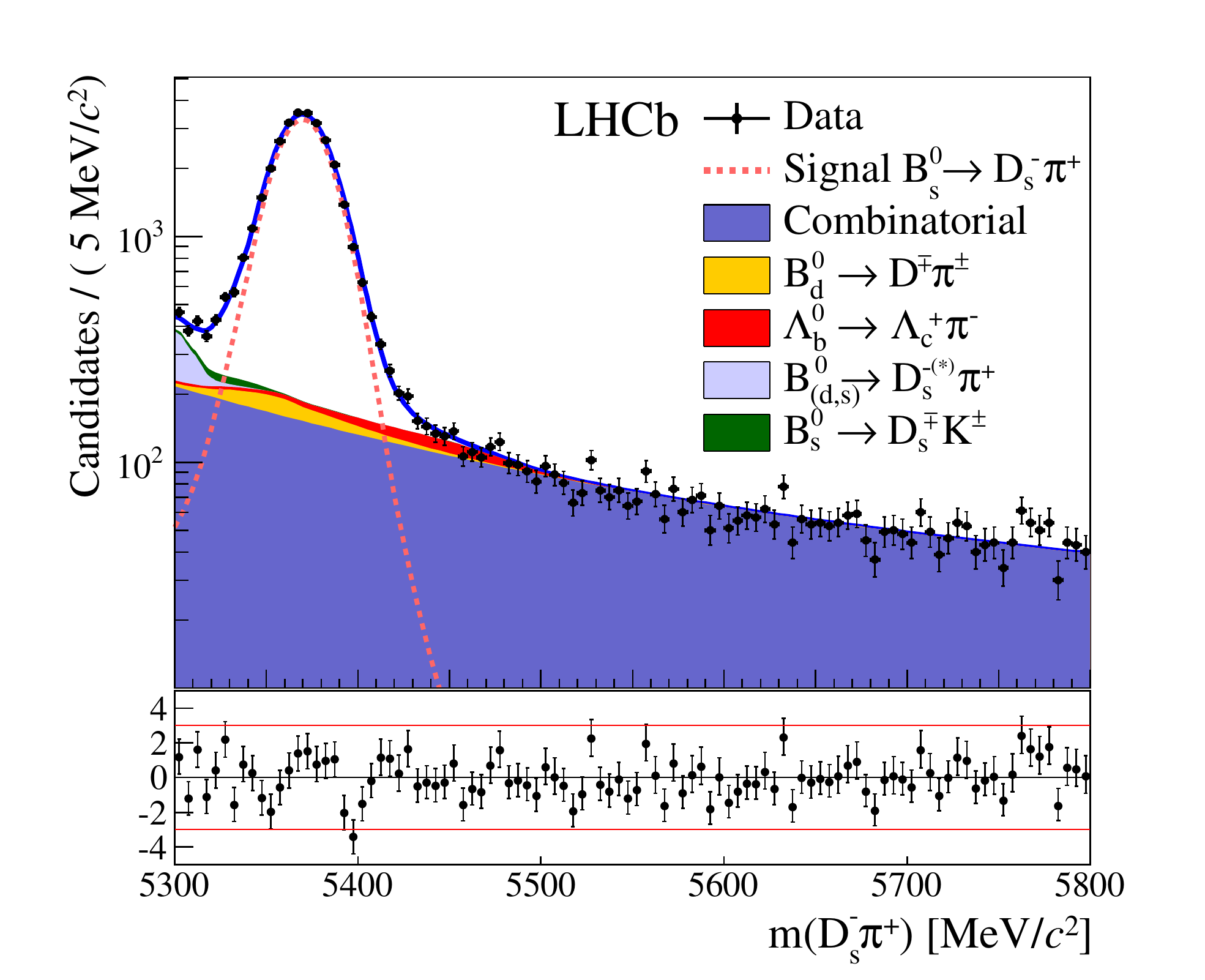}
  \includegraphics[width=.49\textwidth]{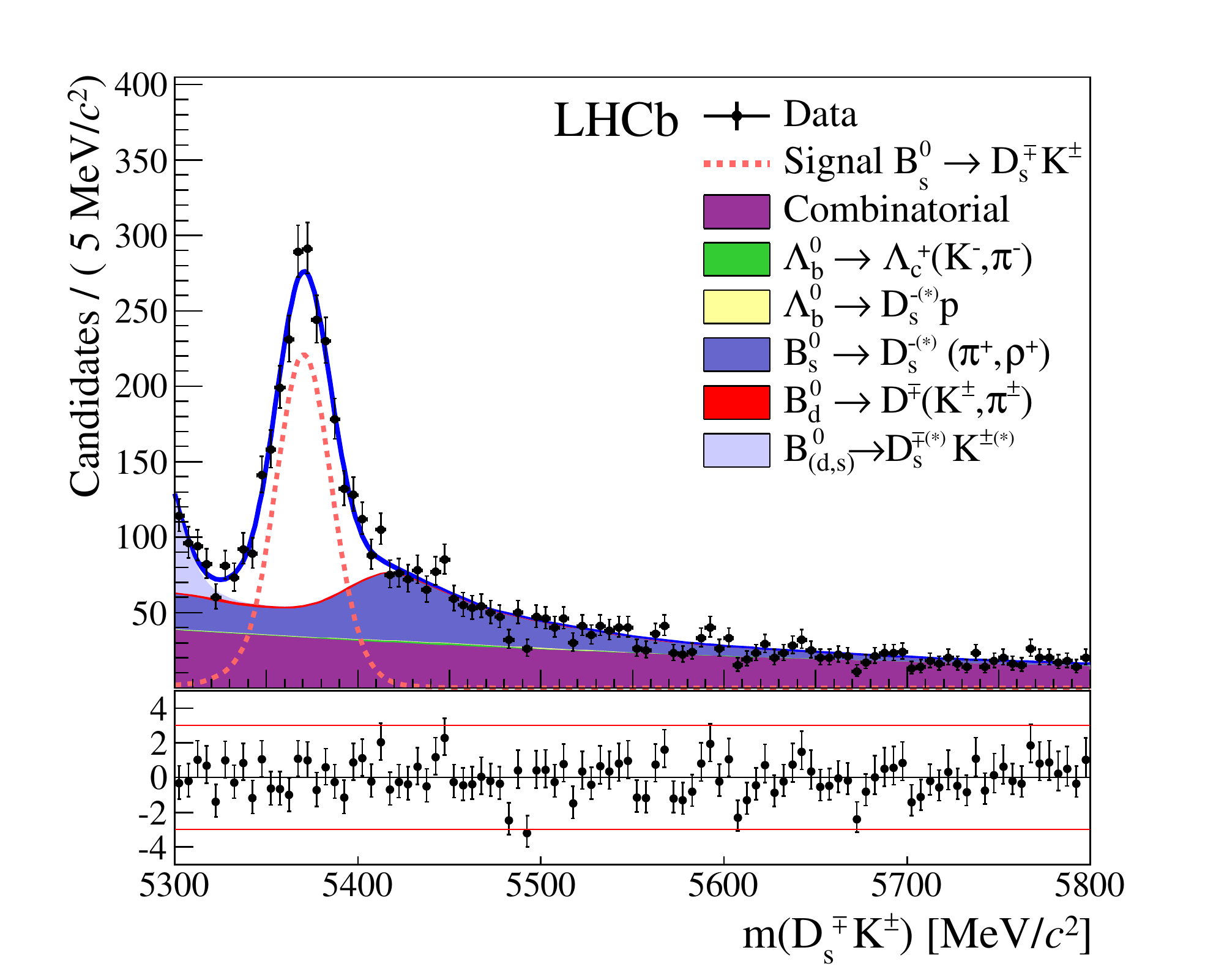}
  \includegraphics[width=.49\textwidth]{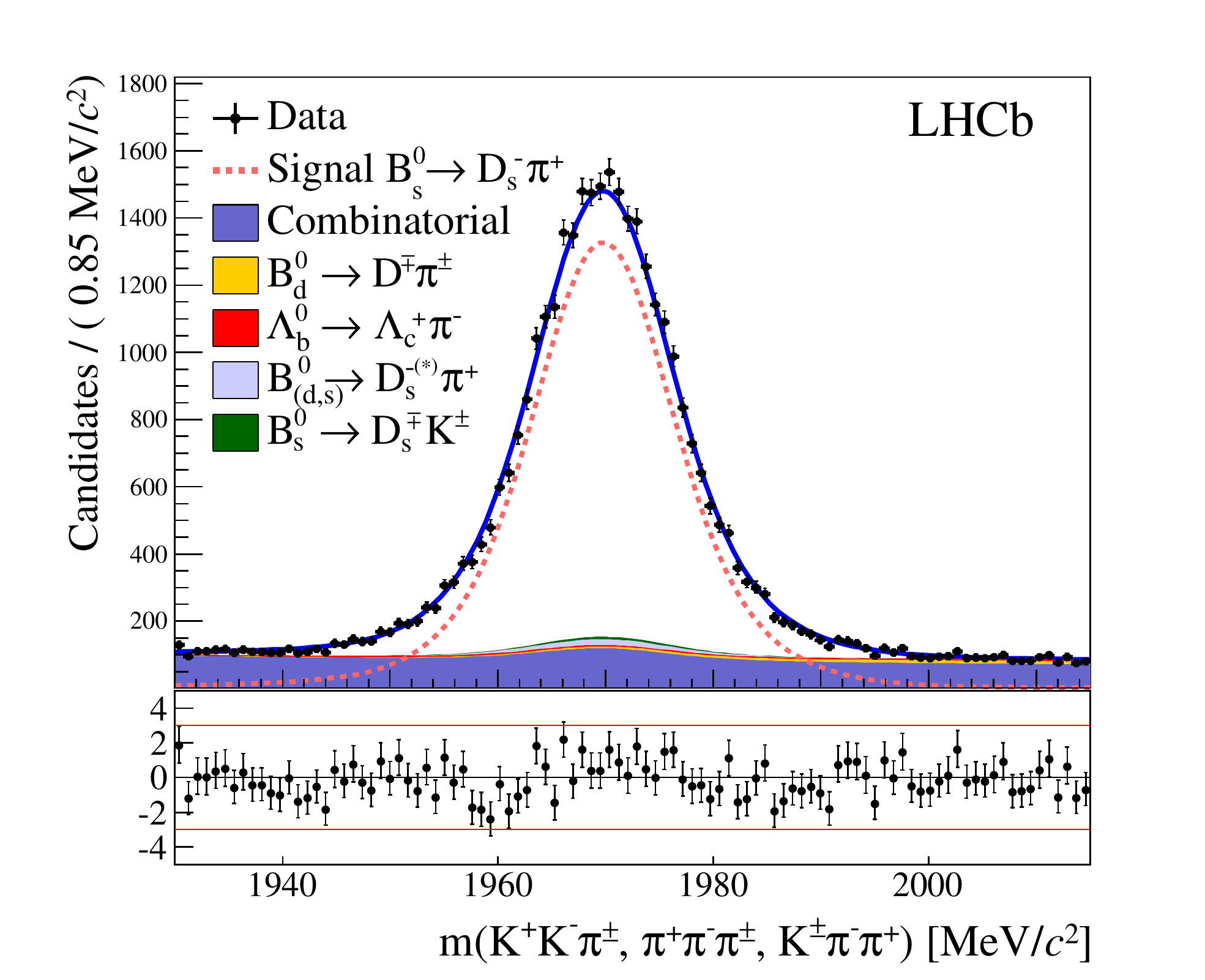}
  \includegraphics[width=.49\textwidth]{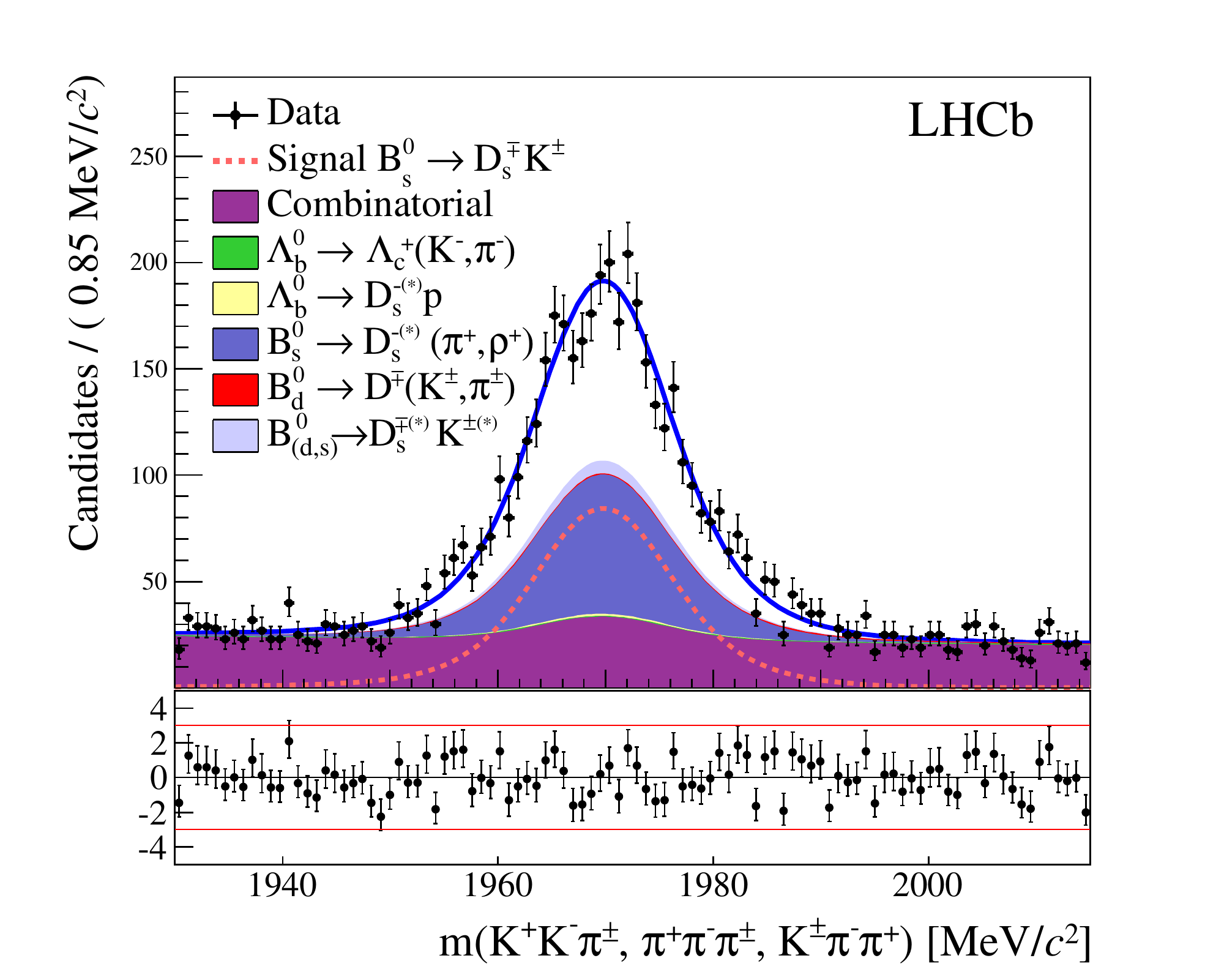}
  \includegraphics[width=.49\textwidth]{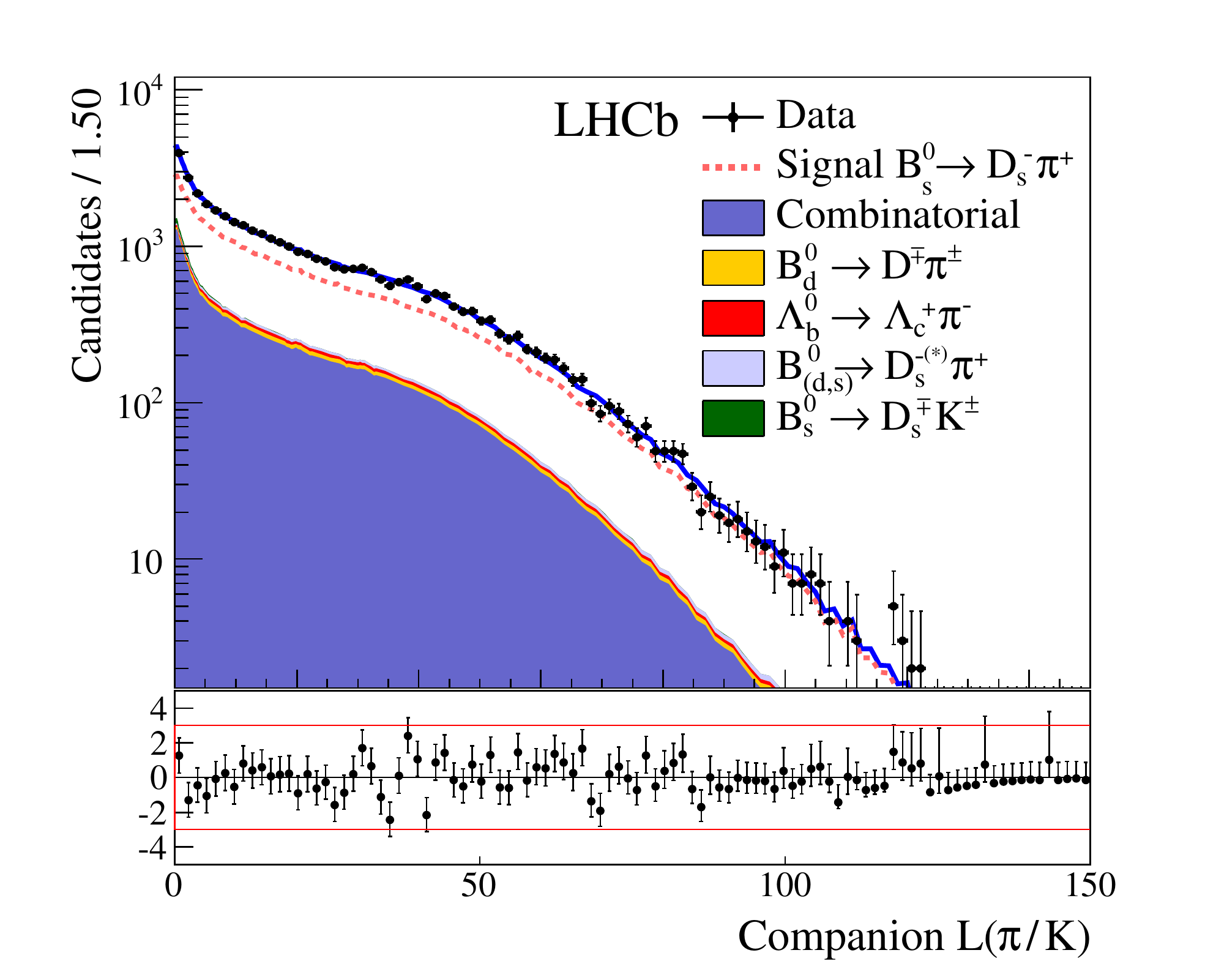}
  \includegraphics[width=.49\textwidth]{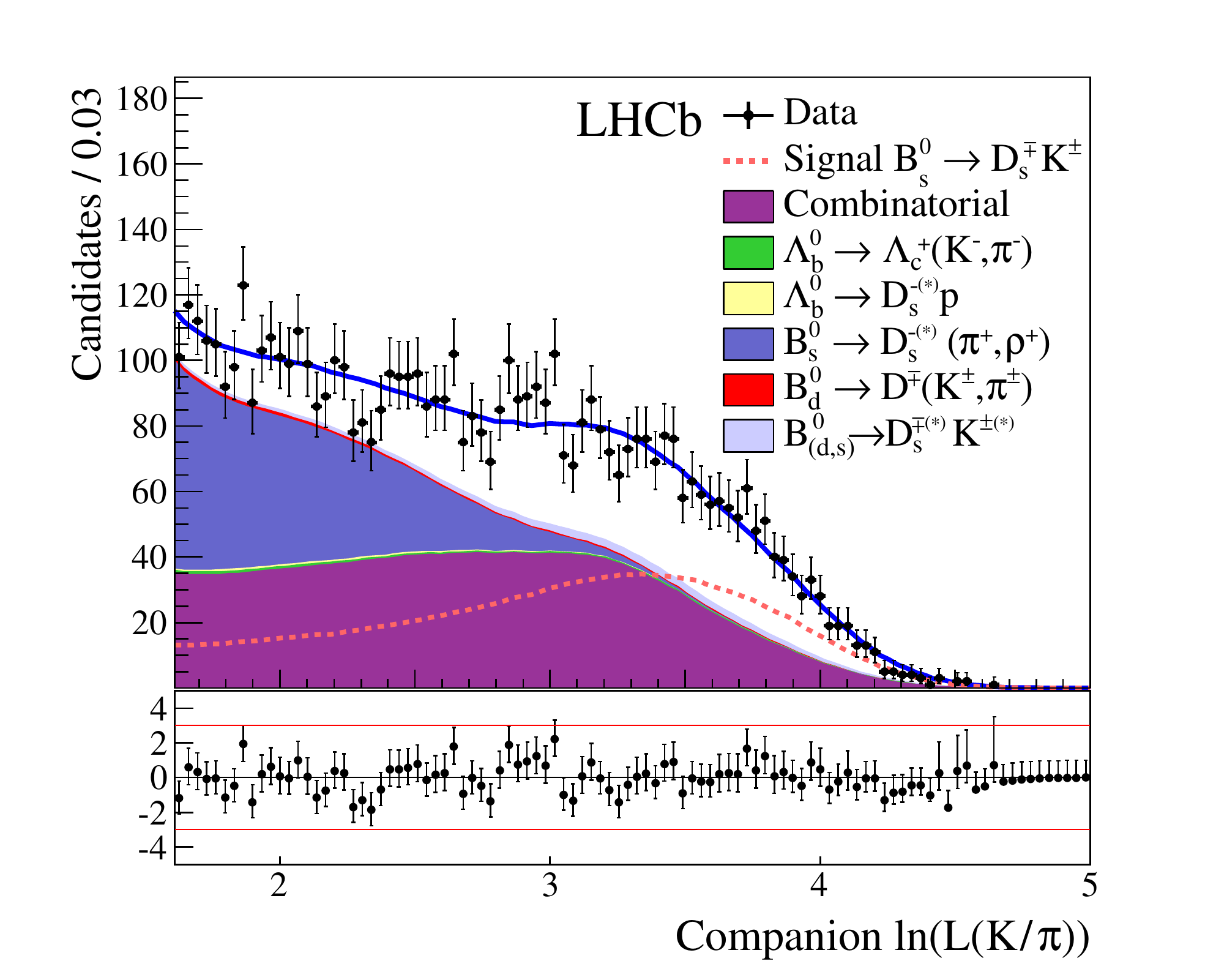}
  \caption{The multivariate fit to the (left) \BsDsPi and (right) \BsDsK candidates for all
    \Dsm decay modes combined. From top to bottom: distributions of candidates
in \Bs mass, \Dsm mass, companion PID log-likelihood difference. The solid, blue, line
represents the sum of the fit components.  }
  \label{fig:massfit-BsDsPiK-All}
\end{figure}

The total PDF for the multivariate fit is built from the product of the signal and background PDFs,
since correlations between the fitting variables are measured to be small in simulation.
These product PDFs are then added for each \Dsm decay mode, and almost all background yields are left free to float. The only exceptions
are those backgrounds whose yield is below $2\%$ of the signal yield. These are \BdDK, \BdDPi, \LbLcK, and \LbLcPi for the \BsDsK
fit, and \BdDPi, \LbLcPi, and \BsDsK for the \BsDsPi fit.
These background yields are fixed from known branching fractions and relative efficiencies measured using simulated events.
The multivariate fit results in a signal yield of $28\,260\pm 180$ \BsDsPi and $1770 \pm 50$ \BsDsK decays,
with an effective purity of $85\%$ for \BsDsPi and $74\%$ for \BsDsK. The multivariate fit is checked for biases
using large
samples of data-like pseudoexperiments, and none is found.
The results of the multivariate fit are shown in Fig.~\ref{fig:massfit-BsDsPiK-All} for both the \BsDsPi and
\BsDsK, summed over all \Dsm decay modes.

\clearpage

\section{Flavour Tagging}
\label{sec:tagging}

At the LHC, $b$ quarks are produced in pairs $b\bar{b}$; one of the two
hadronises to form the signal $B^0_s$, the other $b$ quark hadronises and
decays independently.  The identification of the \Bs initial flavour is
performed by means of two flavour-tagging algorithms which exploit this pair-wise production of $b$ quarks.
The opposite side (OS) tagger determines the flavour of the non-signal
$b$-hadron produced in the proton-proton collision using the charge
of the lepton ($\mu$, $e$) produced in semileptonic $B$ decays, or that of the kaon
from the $b\to c\to s$ decay chain, or the charge of the inclusive
secondary vertex reconstructed from $b$-decay products.
The same side kaon (SSK) tagger searches for an additional charged kaon accompanying the fragmentation of the signal \Bs or \Bsb.

Each of these algorithms has an intrinsic mistag rate $\mistag=(\textrm{wrong
tags})/(\textrm{all tags})$ and a tagging efficiency $\etag=(\textrm{tagged
candidates})/(\textrm{all candidates})$. Candidates can be tagged incorrectly due to
tracks from the underlying event, particle misidentifications, or flavour
oscillations of neutral $B$ mesons. The intrinsic mistag \mistag can only be measured in
flavour-specific, self-tagging final states.

The tagging algorithms predict for each \Bs candidate an estimate $\eta$ of
the mistag probability, which should closely follow the intrinsic mistag
\mistag. This estimate $\eta$ is obtained by using a neural network trained on
simulated events whose inputs are the kinematic, geometric, and PID properties
of the tagging particle(s).

The estimated mistag $\eta$ is treated as a per-candidate variable, thus
adding an observable to the fit. Due to variations in the properties of tagging
tracks for different channels, the predicted mistag probability $\eta$ is usually not exactly the
(true) mistag rate \mistag, which requires $\eta$ to be calibrated using
flavour specific, and therefore self-tagging, decays.
The statistical uncertainty on \Cpar, \Spar, and \Sbpar scales with
$1/\sqrt{\effeff}$, defined as $\effeff = \etag ( 1 - 2 \mistag)^2$.
Therefore, the tagging algorithms are tuned for maximum effective tagging
power \effeff.

\subsection{Tagging calibration}
\label{sec:OS}

The calibration for the OS tagger is performed using
several control channels: $\Bp\to\jpsi\Kp$, $\Bp\to\Dzb\pip$, $\Bd\to D^{*-} \mu^+ \nu_{\mu}$, $\Bd\to\jpsi\Kstarz$ and  \BsDsPi.
This calibration of $\eta$ is done for each control channel using the linear function
\begin{align}
  \label{eq:tagging-calibration}
  \mistag = p_0 + p_1 \cdot (\eta - \langle\eta\rangle)\,,
\end{align}
where the values of $p_0$ and $p_1$ are called calibration parameters, and
$\langle\eta\rangle$ is the mean of the $\eta$ distribution predicted by a
tagger in a specific control channel.
Systematic uncertainties are assigned to account for possible dependences of
the calibration parameters on the final state considered, on the kinematics of
the \Bs candidate and on the event properties.
The corresponding values of the calibration parameters are summarised in Table~\ref{tab:OS-tagging-calibration-all}.
For each control channel the relevant calibration parameters are reported with their statistical and systematic uncertainties.
These are averaged to give the reference values including a systematic uncertainty accounting for kinematic 
differences between different channels. The resulting calibration parameters for the \BsDsK fit
are: $p_0 =  0.3834 \pm 0.0014 \pm 0.0040$ and $p_1 =  0.972 \pm 0.012 \pm 0.035$, where the $p_0$ for each control channel 
needs to be translated to the $\langle\eta\rangle$ of \BsDsPi, the channel
which is most similar to the signal channel \BsDsK. This is achieved by the transformation
$p_0\to p_0+p_1 (\langle\eta\rangle - 0.3813)$ in each control channel.

\begin{table}[bt]
\centering
\caption{Calibration parameters of the combined OS tagger
        extracted from different control channels. In each entry the first uncertainty is statistical and the second systematic.
}
\label{tab:OS-tagging-calibration-all}
\small
\begin{tabular}{lccc}
\hline
Control channel   & $\langle\eta\rangle$ & $p_0 - \langle\eta\rangle$            & $p_1$             \\
\hline
$\Bp\to\jpsi\Kp$                & 0.3919 & 0.0008 $\pm$ 0.0014 $\pm$ 0.0015 & 0.982 $\pm$ 0.017 $\pm$ 0.005 \\
$\Bp\to\Dzb\pip$                 & 0.3836 & 0.0018 $\pm$ 0.0016 $\pm$ 0.0015 & 0.972 $\pm$ 0.017 $\pm$ 0.005 \\
$\Bd\to\jpsi\Kstarz$            & 0.390\phantom{0} & 0.0090  $\pm$ 0.0030  $\pm$ 0.0060  & 0.882 $\pm$ 0.043 $\pm$ 0.039 \\
$\Bd\to D^{*-} \mu^+ \nu_{\mu}$ & 0.3872 & 0.0081 $\pm$ 0.0019 $\pm$ 0.0069 & 0.946 $\pm$ 0.019 $\pm$ 0.061 \\
\BsDsPi                         & 0.3813 & 0.0159 $\pm$ 0.0097 $\pm$ 0.0071 & 1.000 $\pm$ 0.116 $\pm$ 0.047 \\
\hline
Average & 0.3813 & 0.0021 $\pm$ 0.0014 $\pm$ 0.0040 & 0.972 $\pm$ 0.012 $\pm$ 0.035\\
\hline
\end{tabular}
\end{table}

The SSK algorithm uses a neural network to select fragmentation particles,
giving improved flavour tagging power~\cite{Krocker} with respect to earlier cut-based~\cite{LHCb-CONF-2012-033} algorithms. It is
calibrated using the \BsDsPi channel, resulting in  $\langle\eta\rangle$ = 0.4097,
$p_0 =   0.4244 \pm 0.0086 \pm 0.0071$ and $p_1 =  1.255 \pm 0.140 \pm0.104$,
where the first uncertainty is statistical and second systematic. The systematic uncertainties include the uncertainty
on the decay-time resolution, the \BsDsPi fit model, and the backgrounds in the \BsDsPi fit.

\begin{figure}[tb]
  \centering
  \includegraphics[width=.49\textwidth]{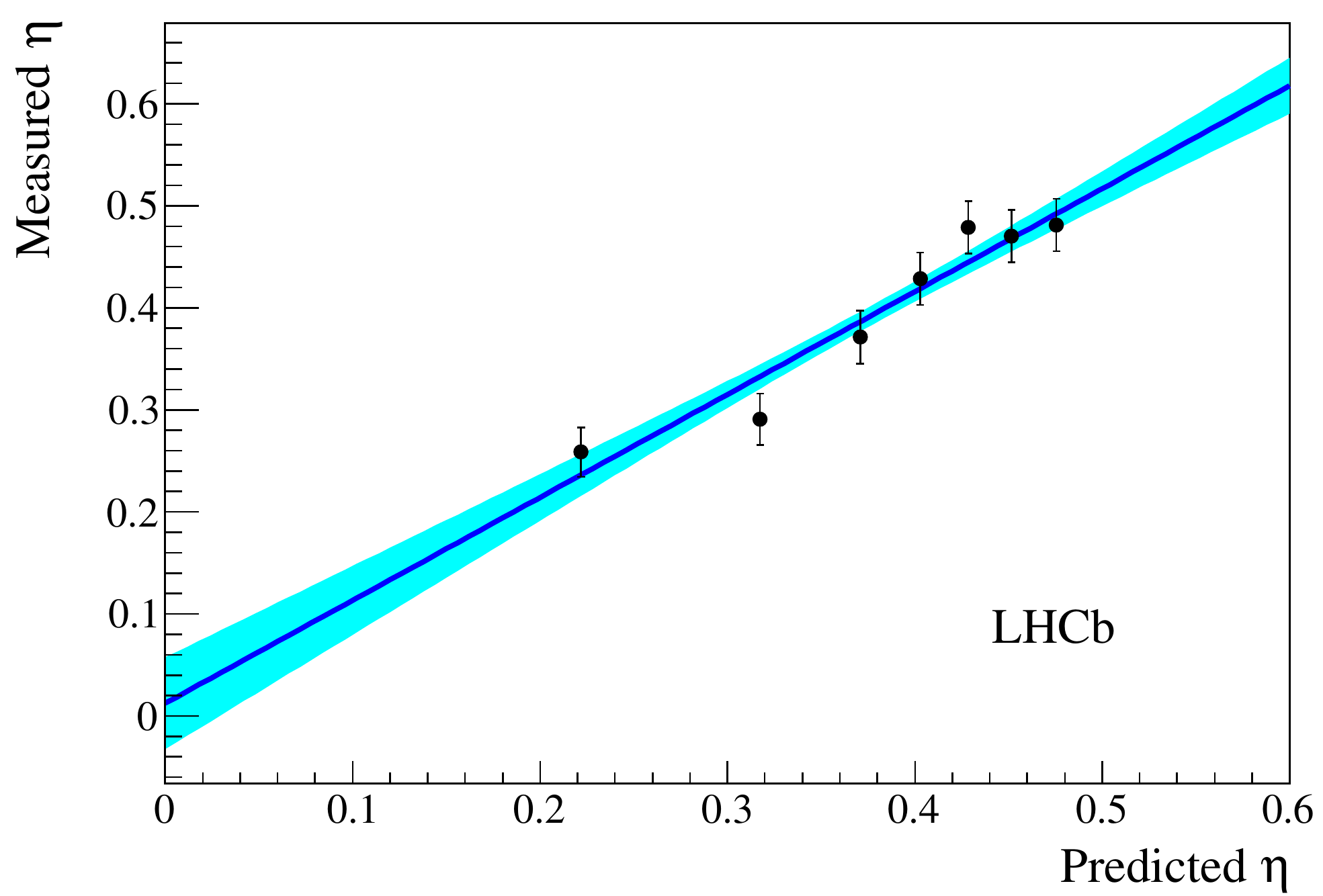}
  \includegraphics[width=.49\textwidth]{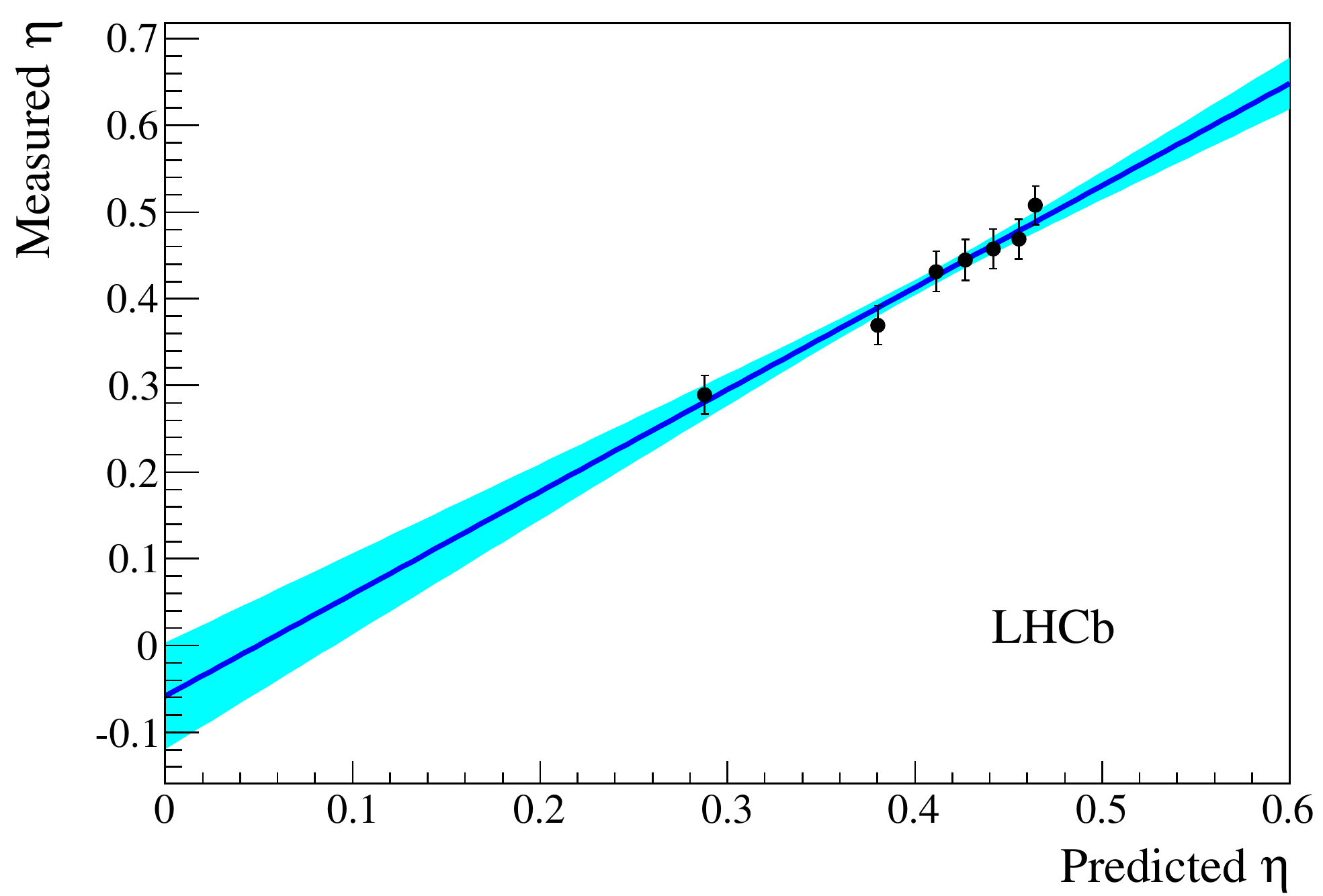}
  \caption{Measured mistag rate against the average predicted mistag rate for the (left) OS and (right) SSK taggers in \BsDsPi decays.
    The error bars represent only the statistical uncertainties.
    The solid curve is the linear fit to the data points, the shaded area defines the 68\% confidence level region of the calibration function (statistical only).
    }
  \label{fig:OS-tagging-calibration}
\end{figure}

Figure~\ref{fig:OS-tagging-calibration} shows the measured mistag probability as
a function of the mean predicted mistag probability in \BsDsPi decays for the OS and SSK taggers. The data points show a linear correlation
corresponding to the functional form in Eq.~\ref{eq:tagging-calibration}.
We additionally validate that the obtained tagging calibration parameters can be used in \BsDsK decays by comparing them
for \BsDsK and \BsDsPi in simulated events; we find excellent agreement between the two.
We also evaluate possible tagging asymmetries between \B and \Bbar mesons 
for the OS and SSK taggers by performing the calibrations split by \B meson flavour. The OS tagging asymmetries
are measured using $\Bp\to\jpsi\Kp$ decays, while the SSK tagging asymmetries are measured using
prompt $D^{\pm}_{s}$ mesons whose \pt distribution has been weighted to match the \BsDsPi signal.
The resulting initial flavour asymmetries for $p_{0}$, $p_{1}$ and \etag are taken into account in the decay-time fit.

\subsection{Combination of OS and SSK taggers}
\label{sec:taggingCombination}

Since the SSK and OS taggers rely on different physical processes they are largely independent, with a correlation measured as negligible.
The tagged candidates are therefore split into three different samples depending on the
tagging decision: events only tagged by the OS tagger (OS-only), those only tagged by the SSK tagger (SSK-only),
and those tagged by both the OS and SSK taggers (OS-SSK).
For the candidates that have decisions from both taggers a combination is performed using the calibrated mistag probabilities.
The combined tagging decision and calibrated mistag rate are used in the final time-dependent fit,
where the calibration parameters are constrained using the combination of their associated statistical and systematic uncertainties.
The tagging performances, as well as the effective tagging power,
for the three sub-samples and their combination as measured using \BsDsPi events
are reported in Table~\ref{tab:tagging-performances}.

\begin{table}[bt]
  \caption{Flavour tagging performance for the three different tagging categories
  for \BsDsPi candidates.}
  \label{tab:tagging-performances}
  \centering
  \begin{tabular}{lcc}
    \hline
    Event type & \etag  [\%]                       & \effeff [\%]          \\
    \hline
    OS-only & 19.80 $\pm$ 0.23 & 1.61 $\pm$ 0.03 $\pm$ 0.08                 \\
    SSK-only & 28.85 $\pm$ 0.27 & 1.31 $\pm$ 0.22 $\pm$ 0.17 \\
    OS-SSK &18.88 $\pm$ 0.23 & 2.15 $\pm$ 0.05 $\pm$ 0.09 \\
    \hline
    Total        &            67.53       &     5.07                                       \\
    \hline
 \end{tabular}
\end{table}

\subsection{Mistag distributions}

Because the fit uses the per-candidate mistag prediction, it is necessary to model the distribution
of this observable for each event category (SS-only, OS-only, OS-SSK for the signal and each background category).
The mistag probability distributions for all \Bs decay modes, whether signal or background,
are obtained using \sWeighted \BsDsPi events. 
The mistag probability distributions for combinatorial
background events are obtained from the upper \Bs mass sideband in \BsDsPi decays. For \Bd and \Lb backgrounds the mistag
distributions are obtained from \sWeighted \BdDPi events. For the SSK tagger this is justified by the fact that these
backgrounds differ by only one spectator quark and should therefore have similar properties with respect to the fragmentation of the $s\overline{s}$ pair.
For the OS tagger, the predicted mistag distributions mainly depend on the kinematic properties of the $B$ candidate, which
are similar for \Bd and \Lb backgrounds.

\section{Decay-time resolution and acceptance}
% --------------------------------------------
\label{sec:timeresandacc}
The decay-time resolution of the detector must be accounted for because of the fast
\Bs--\Bsb oscillations. Any mismodelling of the resolution
function also potentially biases and affects the precision of the
time-dependent \CP violation observables.
The signal decay-time PDF is convolved with a resolution function that has a different width
for each candidate, making use of the per-candidate decay-time uncertainty
estimated by the decay-time kinematic fit.
This approach requires the per-candidate decay-time uncertainty to be calibrated.
The calibration is performed using prompt \Dsm mesons combined with a random track and
kinematically weighted to give a sample of ``fake \Bs'' candidates, which have a
true lifetime of zero. From the spread of the observed decay times, a scale factor to the 
estimated decay time resolution is found to be $1.37\pm0.10$ \cite{LHCb-PAPER-2013-006}. Here the
uncertainty is dominated by the systematic uncertainty
on the similarity between the kinematically weighted ``fake \Bs'' candidates and the signal.
As with the per-candidate mistag, the distribution of per-candidate decay-time uncertainties is
modelled for the signal and each type of background. For the signal these distributions
are taken from \sWeighted data, while for the combinatorial background they are
taken from the \Bs mass sidebands. For other backgrounds, the decay-time error
distributions are obtained from simulated events, which are weighted for the
data-simulation differences found in \BsDsPi signal events.

In the case of background candidates which are either partially reconstructed or in which a particle is misidentified,
the decay-time is incorrectly estimated because either the measured mass of the background candidate, the measured
momentum, or both, are systematically misreconstructed. For example, in the case of \BsDsPi as a background to \BsDsK,
the momentum measurement is unbiased, while the reconstructed mass is systematically above the true mass,
leading to a systematic increase in the reconstructed decay-time. This effect causes
an additional non-Gaussian smearing of the decay-time distribution,
which is accounted for in the decay time resolution by nonparametric PDFs obtained
from simulated events, referred to as $k$-factor templates.

\begin{figure}[tb]
 \centering
 \includegraphics[width=.48\textwidth]{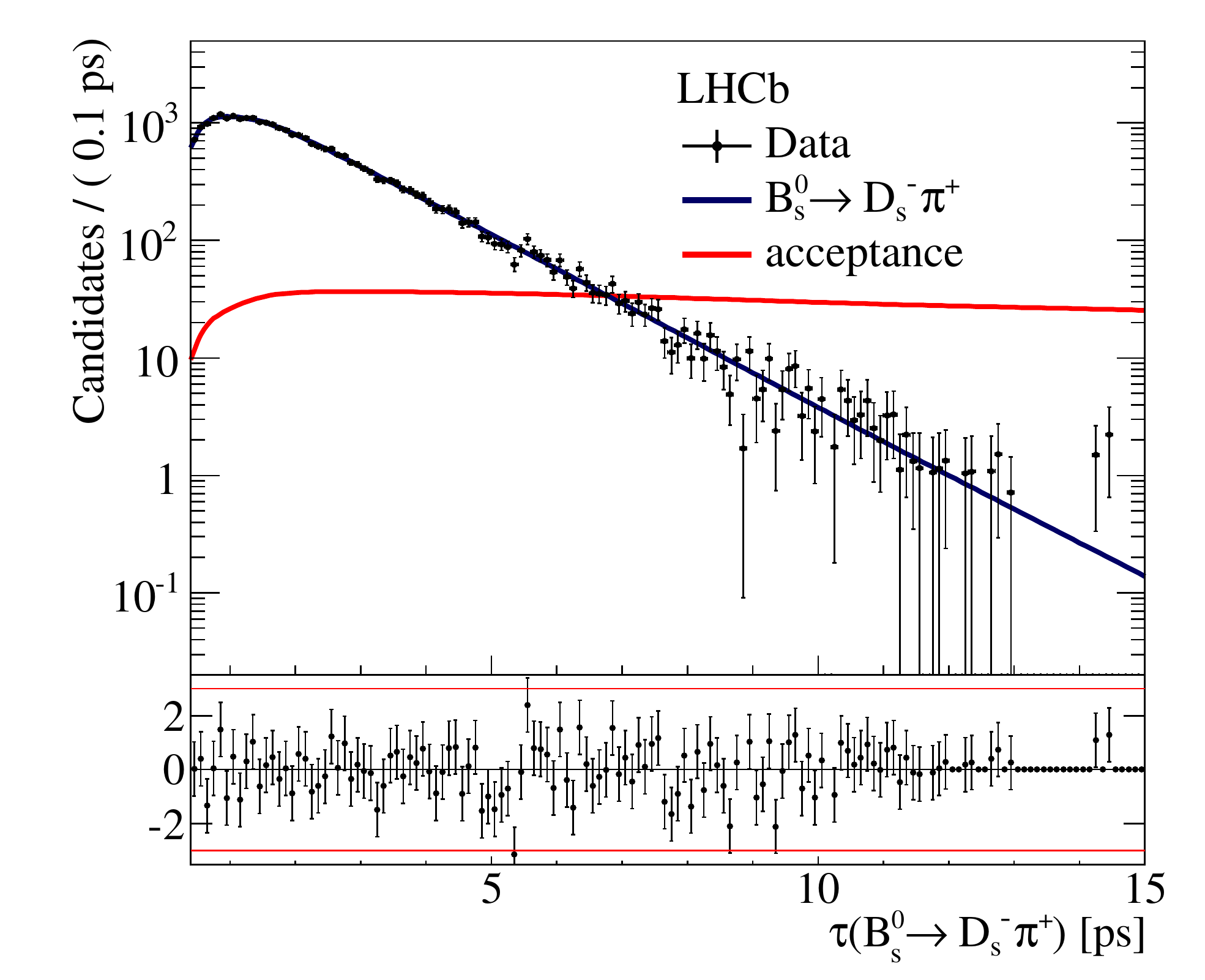}
 \caption{Result of the \sfit to the decay-time distribution of \BsDsPi candidates, which is used to measure the decay-time acceptance in \BsDsK decays.
The solid curve is measured decay-time acceptance.}
\label{fig:timesfit_bsdspi}
\end{figure}

The decay-time acceptance of \BsDsK candidates cannot be floated because
its shape is heavily correlated
with the \CP observables. In particular the upper decay-time acceptance is correlated
with \Dpar and \Dbpar. However, in the case of \BsDsPi, the acceptance can be measured 
by fixing \gs and floating the acceptance parameters. 
The decay-time acceptance in the \BsDsK fit is fixed
to that found in the \BsDsPi data fit, corrected by the acceptance ratio in the two channels in simulated signal events.
These simulated events have been weighted in the manner described in Sec.~\ref{sec:sigbacklineshapes}.
In all cases, the acceptance is described using 
segments of smooth polynomial functions (``splines''), which can be implemented in an analytic way in the decay-time fit \cite{Karbach:ComplxErrFunc}. 
The spline boundaries (``knots'') were chosen in an ad hoc fashion to model reliably
the features of the acceptance shape, and placed at $0.5$, $1.0$, $1.5$, $2.0$, $3.0$, $12.0\ps$.
Doubling the number of knots results in negligible changes to the nominal fit result. The decay-time fit to the \BsDsPi data
is an \sfit using the signal PDF from Sec.~\ref{sec:equations}, with \Spar, \Sbpar, \Dpar, and \Dbpar all fixed
to zero, and the knot magnitudes and \dms floating. 
The measured value of $\dms = 17.772 \pm 0.022 \invps$ (the uncertainty is statistical only) is in excellent agreement with the published LHCb measurement
of $\dms = 17.768 \pm 0.023\pm0.006 \invps$~\cite{LHCb-PAPER-2013-006}. The time
fit to the \BsDsPi data together with the measured decay-time acceptance is shown in Fig.~\ref{fig:timesfit_bsdspi}.

\section{Decay-time fit to $\mathbf{\BsDsK}$}
\label{sec:timefit}

As described previously, two decay-time fitters are used: in one all signal and background time distributions are
described (\hspace{-0.125em}\cfit), and in a second the background is statistically subtracted
using the \sPlot technique~\cite{Pivk:2004ty} where only the signal time distributions
are described (\hspace{-0.125em}\sfit). In both cases an unbinned maximum likelihood fit is performed to the \CP observables defined
in Eq.~\ref{eq:asymm_obs}, and the signal decay-time PDF is identical in the two fitters.
Both the signal and background PDFs are described in the remainder of this section,
but it is important to bear in mind that none of the information
about the background PDFs or fixed background parameters is relevant for the \sfit. 
When performing the fits to the decay-time distribution, the following
parameters are fixed from independent measurements~\cite{LHCb-PAPER-2013-002,LHCb-PAPER-2014-003,PDG}:
\begin{align*}
\gs          &= 0.661 \pm 0.007\invps\,,&\dgs     &= \ifthenelse{\boolean{uselhcbcpconvention}}{}{-}0.106 \pm 0.013\invps\,,&&\rho(\gs,\dgs) = -0.39\,,\\
\Gamma_{\Lb} &= 0.676 \pm 0.006\invps\,,&\Gamma_d &= 0.658 \pm 0.003\invps                                               \,,&&\dms = 17.768 \pm 0.024\invps\,.
\end{align*}
Here $\rho(\gs,\dgs)$ is the correlation between these two measurements, $\Gamma_{\Lb}$ is the decay-width of the \Lb baryon,
$\Gamma_d$ is the \Bd decay width, and \dms is the \Bs oscillation frequency.

The signal production asymmetry is fixed to zero because the fast \Bs oscillations wash out any initial asymmetry
and make its effect on the \CP observables negligible. The signal
detection asymmetry is fixed to $(1.0\pm0.5)\%$, with the sign convention in which
positive detection asymmetries correspond to a higher efficiency to reconstruct positive kaons~\cite{LHCB-PAPER-2014-013,LHCB-PAPER-2013-053}.
The background production and detection asymmetries are floated within constraints of $\pm 1\%$
for \Bs and \Bd decays, and $\pm 3\%$ for \Lb decays.

The signal and background mistag
and decay-time uncertainty distributions, including $k$-factors, are modelled by
kernel templates as described in Sec.~\ref{sec:tagging}~and~\ref{sec:timeresandacc}.
The tagging calibration parameters are constrained to the values obtained from the control
channels for all \Bs decay modes, except for \Bd and \Lb decays where the calibration parameters of the SSK tagger are fixed to $p_0=0.5$, $p_1 = 0$.
All modes use the same spline-based decay-time acceptance function described in Sec.~\ref{sec:timeresandacc}.

The backgrounds from \Bs decay modes are all flavour-specific, and are modelled by the decay-time
PDF used for \BsDsPi decays convolved with the appropriate decay-time resolution and $k$-factors model
for the given background. The backgrounds from \Lb decay modes are all described by a single exponential
convolved with the appropriate decay-time resolution and $k$-factor models. The \BdDK background is flavour
specific and is described with the same PDF as \BsDsPi, except with \dmd instead of \dms in the oscillating terms,
$\Gamma_d$ instead of \gs and the appropriate decay-time resolution and $k$-factor KEYS templates.
The \BdDPi background, on the other hand, is not a flavour specific decay,
and is itself sensitive to \CP violation as discussed in Sec.~\ref{sec:intro}. Its decay-time PDF therefore
includes nonzero \Spar and \Sbpar terms which are constrained to their world-average values~\cite{PDG}.
The decay-time PDF of the combinatorial background used in the \cfit is a double exponential function split by the tagging category of the event,
whose parameters are measured using events in the \Bs mass sidebands.

All decay-time PDFs include the effects of flavour tagging, are convolved with a single Gaussian
representing the per-candidate decay-time resolution, and are multiplied by the decay-time acceptance
described in Sec.~\ref{sec:timeresandacc}.
Once the decay-time PDFs are constructed, the \sfit proceeds by fitting the signal PDF to the \sWeighted \BsDsK candidates.
The \cfit, on the other hand, performs a six-dimensional fit to the decay time, decay-time error, predicted mistag, and the three variables
used in the multivariate fit. The \Bs mass range is restricted to $m(\Bs) \in [5320,5420]\mevcc$, and the yields of the different
signal and background components are fixed to those found in this fit range in the multivariate fit. The decay-time range of the fit is
$\tau(\Bs) \in [0.4,15.0]\ps$ in both cases. 

\begin{figure}[t]
 \centering
 \includegraphics[width=.48\textwidth]{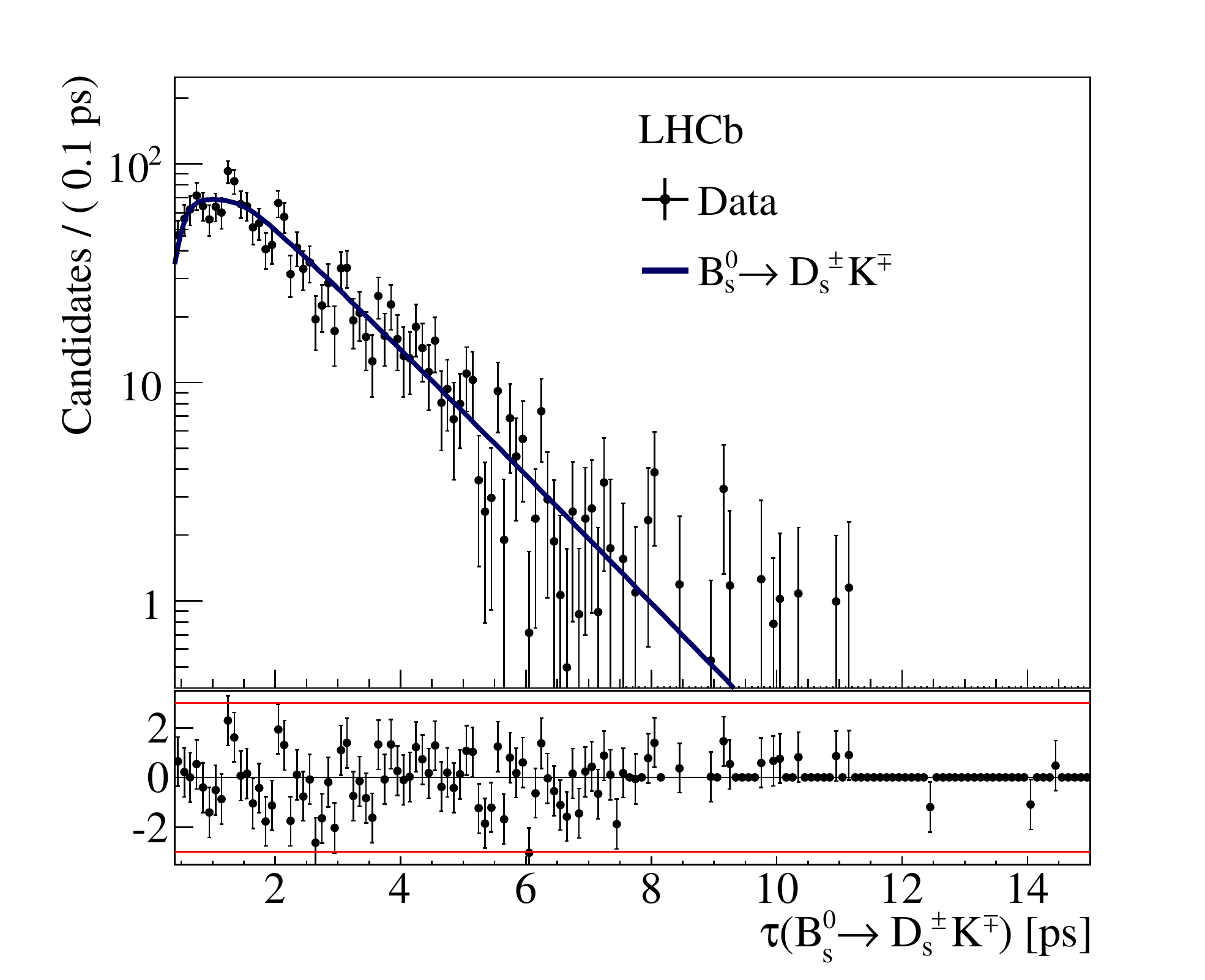}
 \includegraphics[width=.48\textwidth]{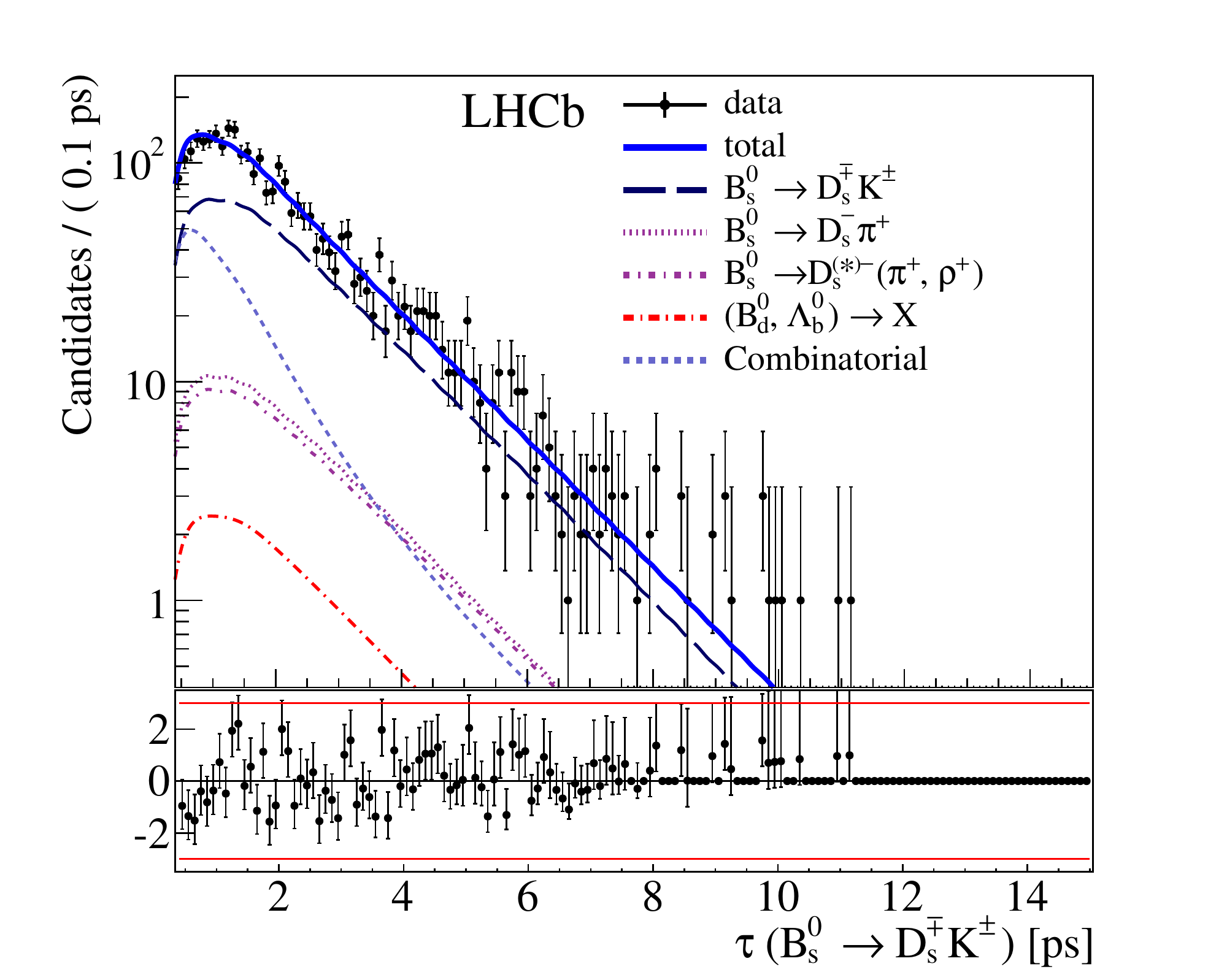}
 \includegraphics[width=.48\textwidth]{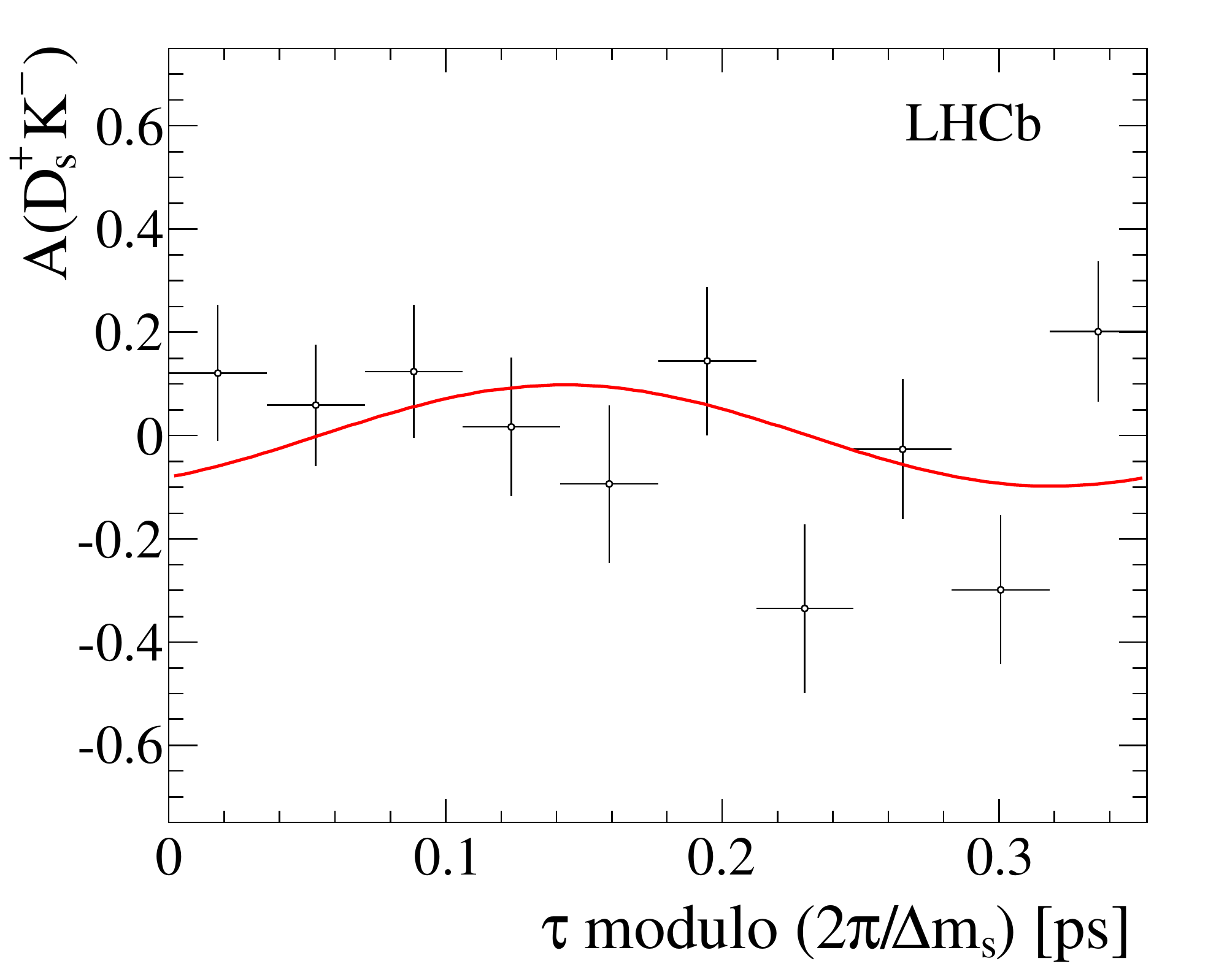}
 \includegraphics[width=.48\textwidth]{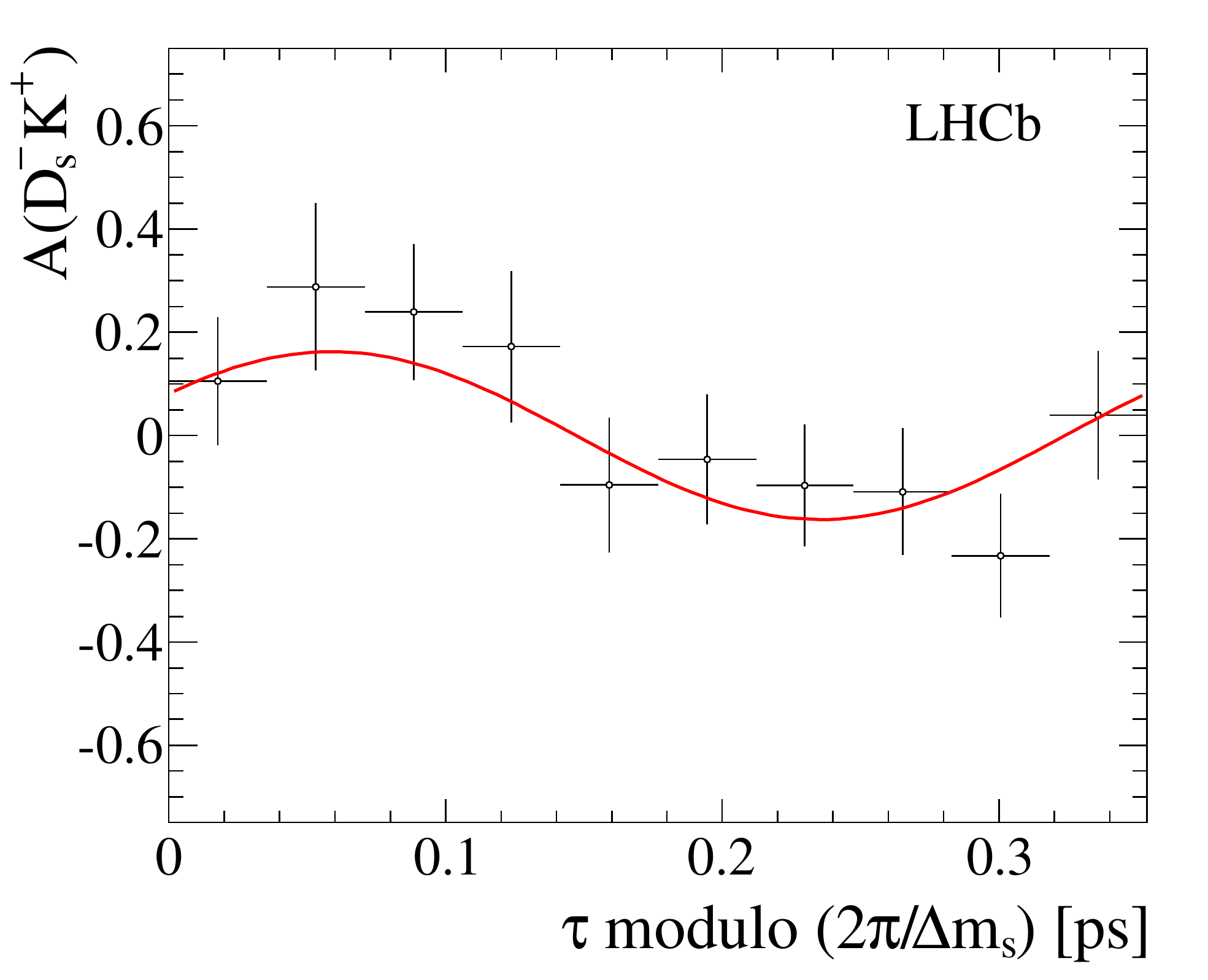}
\caption{Result of the decay-time (top left) \sfit and (top right) \cfit to
     the \BsDsK candidates; the \cfit plot groups \BsDsstPi and \BsDsRho, and
     also groups \BdDK, \BdDPi, \LbLcK, \LbLcPi, \LbDsP, \LbDsstP, and \BdDsK
     together for the sake of clarity.
     The folded asymmetry plots for (bottom left) $\Dsp K^-$, and (bottom right) $\Dsm K^+$ are also shown.}
 \label{fig:timesfit_bsdsk}
\end{figure}

The results of the \cfit and \sfit for the \CP violating observables are given in Table~\ref{tab:timefit_bsdsk}, and
their correlations in Table~\ref{tab:timefit_bsdskcorr}. The fits to the decay-time distribution are shown in Fig.~\ref{fig:timesfit_bsdsk}
together with the folded asymmetry plots for \DspKm and \DsmKp final states. The folded asymmetry plots show the difference in the rates of
\Bs and \Bsb tagged \DspKm and \DsmKp candidates, plotted in slices of $2\pi/\dms$, where the {\it sWeights} obtained with the multivariate fit have been
used to subtract background events. The plotted asymmetry function is drawn using the \sfit central values of the \CP observables,
and is normalised using the expected dilution due to mistag and time resolution.
\clearpage
\begin{table}[b]
\centering
\caption{Fitted values of the \CP observables to the \BsDsK time distribution
for (left) \sfit and (right) \cfit, where the first uncertainty is statistical,
the second is systematic.  All parameters other than the \CP observables are
constrained in the fit. }
\label{tab:timefit_bsdsk}
\begin{tabular}{lcc}
  \hline
  Parameter     & \sfit fitted value & \cfit fitted value\\
  \hline
  \Cpar 	& $\phantom{-}0.52 \pm 0.25 \pm 0.04$  & $\phantom{-}0.53 \pm 0.25 \pm 0.04$ \\  
  \Dpar		& $\phantom{-}0.29 \pm 0.42 \pm 0.17$  & $\phantom{-}0.37 \pm 0.42 \pm 0.20$ \\  
  \Dbpar	& $\phantom{-}0.14 \pm 0.41 \pm 0.18$  & $\phantom{-}0.20 \pm 0.41 \pm 0.20$ \\
  \Spar 	& $          -0.90 \pm 0.31 \pm 0.06$  & $          -1.09 \pm 0.33 \pm 0.08$ \\  
  \Sbpar       	& $          -0.36 \pm 0.34 \pm 0.06$  & $          -0.36 \pm 0.34 \pm 0.08$ \\  
  \hline
\end{tabular}
\vspace{-3mm}
\end{table}
\begin{table}[t]
\centering
\caption{Statistical correlation matrix of the \BsDsK (top) \sfit and (bottom) {\cfit} \CP parameters.
Other fit parameters have negligible correlations with the \CP parameters and are omitted for brevity.
}
\label{tab:timefit_bsdskcorr}
\begin{tabular}{llrrrrr}
  \hline
  \multicolumn{2}{l}{Parameter}         & \Cpar   & \Dpar   & \Dbpar  & \Spar   & \Sbpar\\ 
  \hline
  \sfit
  & $\quad$ \Cpar     & $1.000 $  & $ 0.071$  &$ 0.097 $  & $0.117 $  & $-0.042$  \\
  & $\quad$ \Dpar     & $      $  & $1.000 $  &$ 0.500 $  & $-0.044$  & $-0.003$  \\
  & $\quad$ \Dbpar    & $      $  & $      $  &$ 1.000 $  & $-0.013$  & $- 0.005$  \\
  & $\quad$ \Spar     & $      $  & $      $  &$       $  & $1.000 $  & $ 0.007$  \\
  & $\quad$ \Sbpar    & $      $  & $      $  &$       $  & $      $  & $1.000 $  \\
\hline
  \cfit
  & $\quad$ \Cpar     & $ 1.000$ & $ 0.084$ & $ 0.103$ & $-0.008$ & $-0.045$ \\
  & $\quad$ \Dpar     & $      $ & $ 1.000$ & $ 0.544$ & $-0.117$ & $-0.022$ \\
  & $\quad$ \Dbpar    & $      $ & $      $ & $ 1.000$ & $-0.067$ & $-0.032$ \\
  & $\quad$ \Spar     & $      $ & $      $ & $      $ & $ 1.000$ & $ 0.002$ \\
  & $\quad$ \Sbpar    & $      $ & $      $ & $      $ & $      $ & $ 1.000$ \\
\hline
\end{tabular}
\end{table}

\section{Systematic uncertainties}
% --------------------------------
\label{sec:systematics}
Systematic uncertainties arise from
the fixed parameters \dms, \gs, and \dgs,
and from the limited knowledge of the decay time resolution and acceptance.
These uncertainties are estimated
using large sets of simulated pseudoexperiments, in which the relevant parameters are
varied. The pseudoexperiments are generated 
with the average of the \cfit and \sfit central values reported in Sec.~\ref{sec:timefit}.
They are subsequently processed by the
full data fitting procedure: first the multivariate fit to obtain the \sWeights,
and then the decay time fits. 
The fitted values of the observables are compared between
the nominal fit, where all fixed parameters are kept at their nominal values,
and the systematic fit, where each parameter is varied according to its
systematic uncertainty. A distribution is formed by normalising the resulting
differences to the uncertainties measured in the nominal fit, and the 
mean and width of this distribution are added in quadrature and conservatively assigned as the
systematic uncertainty.
The systematic uncertainty on the acceptance is strongly anti-correlated with that
due to the fixed value of \Gs.
This is because the acceptance parameters are determined from the
fit to \BsDsPi data, where \Gs determines the expected exponential slope,
so that the acceptance parameterises any difference between the observed
and the expected slope.
The systematic pseudoexperiments are also used to compute the
systematic covariance matrix due to each source of uncertainty.

The total systematic covariance matrix is obtained by adding 
the individual covariance matrices. The resulting systematic uncertainties
are shown in Tables~\ref{tab:TotalSystErr} and~\ref{tab:TotalSystCor} relative to the
corresponding statistical uncertainties. 
The contributions from \Gs and \DGs are listed independently for comparison to convey
a feeling for their relative importance. For this comparison, \Gs and \DGs
are treated as uncorrelated systematic effects. When computing the total, however, 
the correlations between these two, as well as between them and the acceptance parameters, are accounted for,
and the full systematic uncertainty which enters into the total is listed as ``acceptance, \gs, \dgs''.
The \cfit contains fixed parameters describing the decay time of the
combinatorial background. These parameters are found to be correlated
to the \CP parameters, and a systematic uncertainty is assigned.

The result is cross-checked by splitting the sample into two subsets according
to the two magnet polarities, the hardware trigger decision, and the BDTG response. 
\begin{table}[h]
\centering
\caption{Systematic errors, relative to the statistical error, for (top) \sfit and (bottom) \cfit.
The daggered contributions (\Gs, \DGs) are given separately for comparison (see text) with the other
uncertainties and are not added in quadrature to produce the total.}
\label{tab:TotalSystErr}
\begin{tabular}{llccccc}
  \hline
  \multicolumn{2}{l}{Parameter}     & \Cpar   & \Dpar   & \Dbpar  & \Spar   & \Sbpar \\
  \hline                                                         
  \sfit & $\Delta m_s$              & $0.062$ & $0.013$ & $0.013$ & $0.104$ & $0.100$\\
  & scale factor                    & $0.104$ & $0.004$ & $0.004$ & $0.092$ & $0.096$\\
  & $\DGs^\dagger$                  & $0.007$ & $0.261$ & $0.286$ & $0.007$ & $0.007$\\
  & $\Gs^\dagger$                   & $0.043$ & $0.384$ & $0.385$ & $0.039$ & $0.038$\\
  & acceptance, \Gs, \DGs           & $0.043$ & $0.427$ & $0.437$ & $0.039$ & $0.038$\\
  & sample splits                   & $0.124$ & $0.000$ & $0.000$ & $0.072$ & $0.071$\\
  & total                           & $0.179$ & $0.427$ & $0.437$ & $0.161$ & $0.160$\\
\hline                                                           
  \cfit & $\Delta m_s$              & $0.068$ & $0.014$ & $0.011$ & $0.131$ & $0.126$\\
  & scale factor                    & $0.131$ & $0.004$ & $0.004$ & $0.101$ & $0.103$\\
  & $\DGs^\dagger$                  & $0.008$ & $0.265$ & $0.274$ & $0.009$ & $0.008$\\
  & $\Gs^\dagger$                   & $0.049$ & $0.395$ & $0.394$ & $0.048$ & $0.042$\\
  & acceptance, \Gs, \DGs           & $0.050$ & $0.461$ & $0.464$ & $0.050$ & $0.043$\\
  & comb.~bkg.~lifetime             & $0.016$ & $0.069$ & $0.072$ & $0.015$ & $0.005$\\
  & sample splits                   & $0.102$ & $0.000$ & $0.000$ & $0.156$ & $0.151$\\
  & total                           & $0.187$ & $0.466$ & $0.470$ & $0.234$ & $0.226$\\
\hline
\end{tabular}
\end{table}
\begin{table}[h]
  \centering
  \caption{Systematic uncertainty correlations for (top) \sfit and (bottom) \cfit.}
  \label{tab:TotalSystCor}
  \begin{tabular}{llrrrrr}
    \hline
    \multicolumn{2}{l}{Parameter} & \Cpar   & \Dpar   & \Dbpar  & \Spar  & \Sbpar \\
    \hline
    \sfit & \Cpar  & $ 1.00$ & $ 0.18$ & $ 0.18$ & $-0.04$ & $-0.04$\\
          & \Dpar  & $     $ & $ 1.00$ & $ 0.95$ & $-0.17$ & $-0.16$\\
          & \Dbpar & $     $ & $     $ & $ 1.00$ & $-0.17$ & $-0.16$\\
          & \Spar  & $     $ & $     $ & $     $ & $ 1.00$ & $ 0.05$\\
          & \Sbpar & $     $ & $     $ & $     $ & $     $ & $ 1.00$\\
    \hline                                         
    \cfit & \Cpar  & $ 1.00$ & $ 0.22$ & $ 0.22$ & $-0.04$ & $-0.03$\\
          & \Dpar  & $     $ & $ 1.00$ & $ 0.96$ & $-0.17$ & $-0.14$\\
          & \Dbpar & $     $ & $     $ & $ 1.00$ & $-0.17$ & $-0.14$\\
          & \Spar  & $     $ & $     $ & $     $ & $ 1.00$ & $ 0.09$\\
          & \Sbpar & $     $ & $     $ & $     $ & $     $ & $ 1.00$\\
    \hline
  \end{tabular}
\end{table}
There is\\
$\,$\\
good agreement between the \cfit and the \sfit in each subsample. However, when the sample is
split by BDTG response, the weighted
averages of the subsamples show a small discrepancy
with the nominal fit for \Cpar, \Spar, and \Sbpar, and a corresponding systematic uncertainty is assigned.
In addition, fully simulated signal and background
events are fitted in order to check for systematic effects due to neglecting correlations between the different
variables in the signal and background PDFs. No bias is found.

A potential source of systematic uncertainty is the imperfect
knowledge on the tagging parameters $p_0$ and $p_1$. Their uncertainties
are propagated into the nominal fits by means of Gaussian constraints,
and are therefore included in the statistical error.
A number of other possible systematic effects were studied, but
found to be negligible. These include
possible production and detection asymmetries, and 
missing or imperfectly modelled backgrounds.
Potential systematic effects due to fixed background yields are evaluated by generating
pseudoexperiments with the nominal value for these yields, and fitting back with the yields
fixed to twice their nominal value. No significant bias is observed and no systematic uncertainty assigned.
No systematic uncertainty is attributed to the imperfect knowledge of the
momentum and longitudinal scale of the detector since both effects are
taken into account by the systematic uncertainty in \dms.

Both the \cfit and \sfit are found to be
unbiased through studies of large ensembles of pseudoexperiments generated at the best-fit point in data.
In addition, differences between the \cfit and \sfit are evaluated from the
distributions of the per-pseudoexperiment differences of the fitted values.
Both fitters return compatible results.
Indeed, an important result of this
analysis is that the \sfit technique has been successfully used in an environment
with such a large number of variables, parameters and categories. The \sfit technique was able to 
perform an accurate subtraction of a variety of time-dependent backgrounds in a multidimensional fit,
including different oscillation frequencies, different tagging behaviours,
and backgrounds with modified decay-time distributions due to misreconstructed particles.

\section{Interpretation}
\label{sec:interpretation}
\begin{figure}[b]
\centering
\includegraphics[width=.60\textwidth]{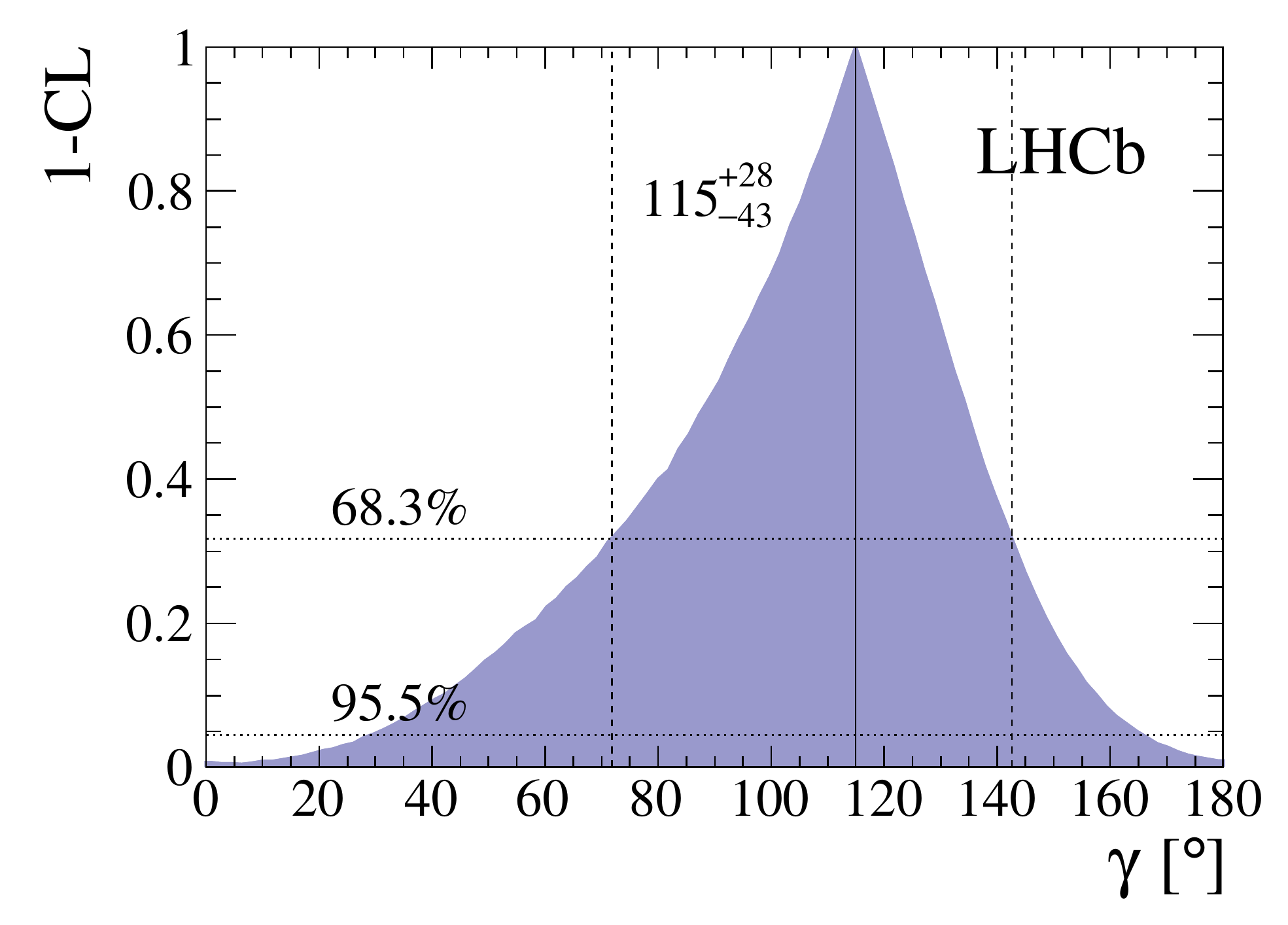}
\includegraphics[width=.48\textwidth]{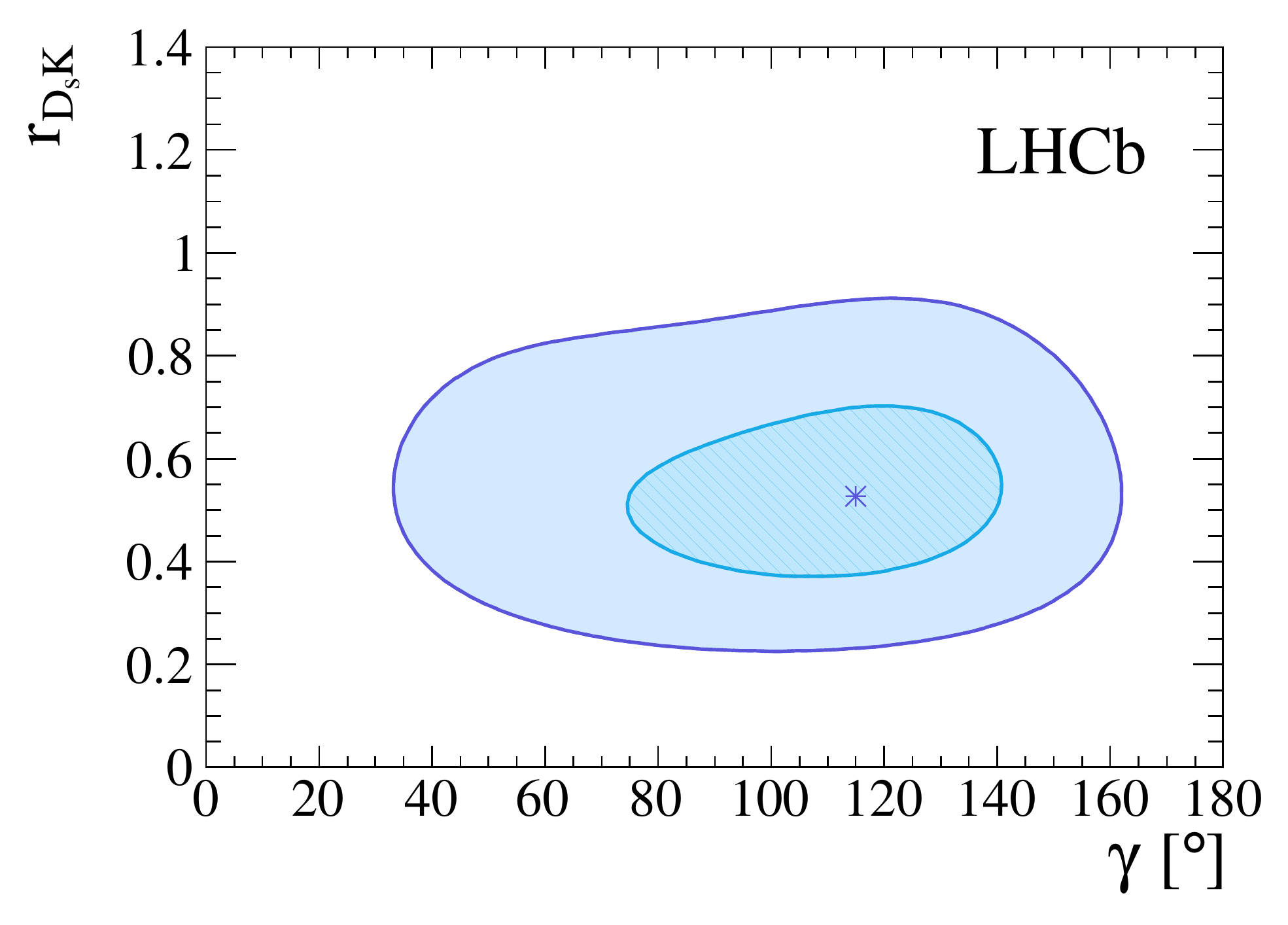}
\includegraphics[width=.48\textwidth]{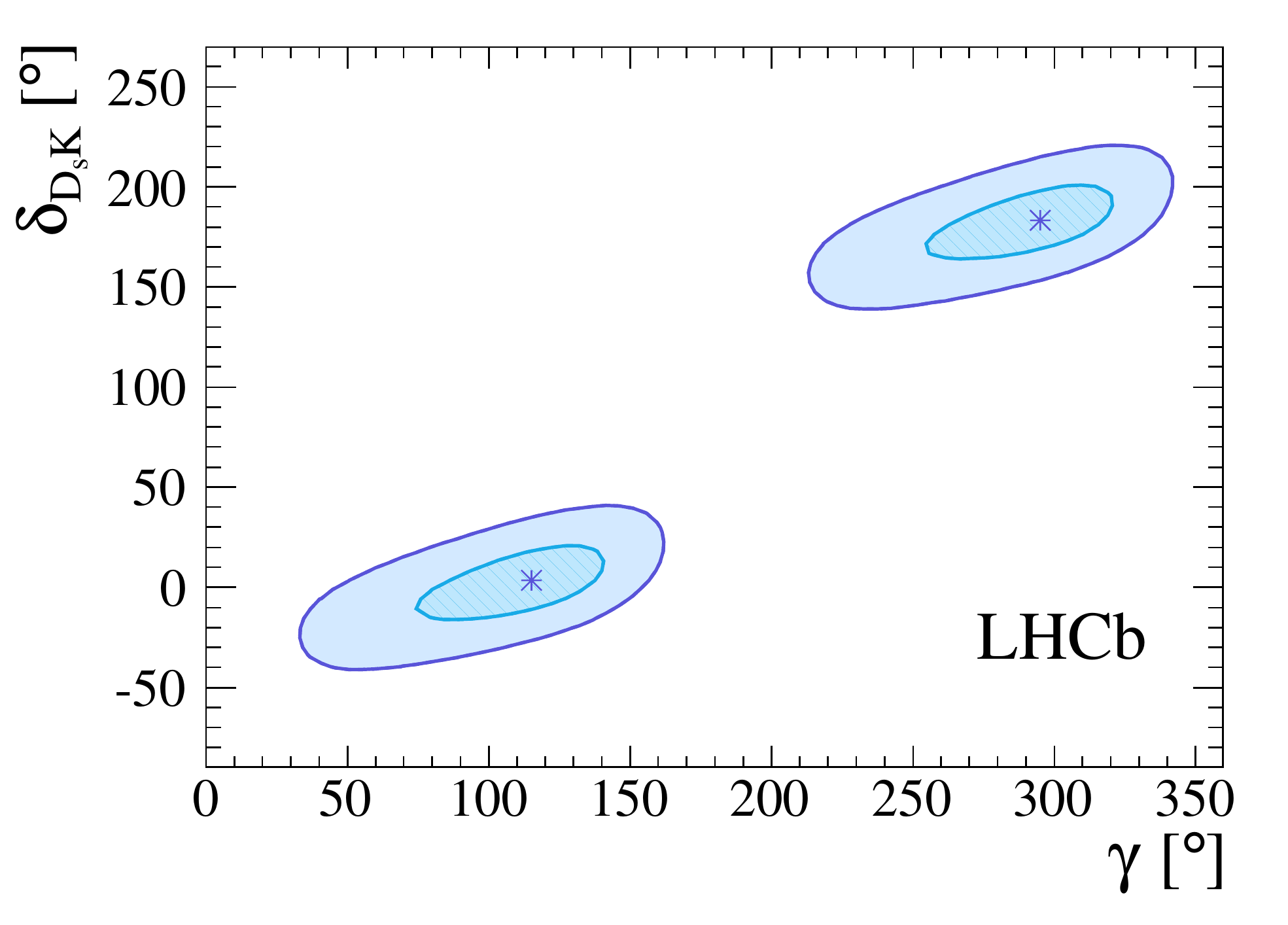}
\caption{\small Graph showing \omcl for \g, together with the central value
and the 68.3\% CL interval as obtained from the frequentist method described
in the text (top). Profile likelihood contours of
\rdsk vs.~\g (bottom left), and
\strong vs.~\g (bottom right).
The contours are the $1\sigma$ ($2\sigma$)
profile likelihood contours, where $\Delta\chi^2=1$ $(\Delta\chi^2=4)$,
corresponding to 39\% CL (86\% CL) in Gaussian approximation.
The markers denote the best-fit values.}
\label{fig:interpretation_gamma_ldsk_ddsk}
\end{figure}
The measurement of the \CP-sensitive parameters is interpreted in terms of
\weak and subsequently \g. 
For this purpose we have arbitrarily chosen the \cfit as the nominal fit result.
The strategy is to maximise the following likelihood
\begin{equation}
  \label{eq:gLikelihood}
  \mathcal L(\vec \alpha) = \exp\left( -\frac 1 2 \left( \vec A(\vec \alpha) - \vec A_{\text{obs}} \right)^T V^{-1} 
  \left( \vec A(\vec \alpha) - \vec A_{\text{obs}} \right) \right)\,,
\end{equation}
where $\vec \alpha = (\g,\phis,\rdsk,\strong)$ is the vector of the physics
parameters, $\vec A$ is the vector of observables expressed through
Eqs.~\ref{eq:truth}, $\vec A_{\text{obs}}$ is the vector of the measured
\CP violating observables and $V$ is the experimental (statistical and systematic)
covariance matrix. Confidence intervals are computed by evaluating the test
statistic 
\mbox{$\Delta\chi^2 \equiv \chi^2(\vec{\alpha}'_{\min}) - \chi^2(\vec{\alpha}_{\min})$},
where $\chi^2(\vec{\alpha}) = -2 \ln \mathcal{L}(\vec{\alpha})$, in a frequentist
way following Ref.~\cite{LHCb-PAPER-2013-020}. Here, $\vec{\alpha}_{\min}$
denotes the global maximum of Eq.~\ref{eq:gLikelihood}, and $\vec{\alpha}'_{\min}$
is the conditional maximum when the parameter of interest is fixed to the tested value.
The value of $\beta_s$ is constrained to the \lhcb measurement from $\Bs\to\jpsi K^+K^-$ and $\Bs\to\jpsi\pi^+\pi^-$ decays, 
$\phis = 0.01 \pm 0.07\stat\pm 0.01\syst\rad$~\cite{LHCb-PAPER-2013-002}.
Neglecting penguin pollution and assuming no BSM contribution in these decays, $\phis = -2\beta_s$.
The resulting confidence intervals are, at 68\% CL,
\begin{align*}
\g      &= (115_{-43}^{+28})^\circ\,,\\
    \strong &= (  3_{-20}^{+19})^\circ\,,\\
\rdsk   &= 0.53_{-0.16}^{+0.17}\,,
\end{align*}
where the intervals for the angles are expressed modulo $180^\circ$.
Figure~\ref{fig:interpretation_gamma_ldsk_ddsk} shows the $1-\textrm{CL}$ curve for
\g, and the two-dimensional
contours of the profile likelihood $\mathcal L(\vec{\alpha}'_{\min})$.
The systematic contributions to the uncertainty are quoted separately as
$\g = \left( 115_{-35}^{+26} \stat {}_{-25}^{+8} \syst \pm 4\,(\phis) \right)^\circ$,
assuming the central value to be independent from systematic uncertainties
and taking the difference in squares of the total
and statistical uncertainties.

\section{Conclusion}
\label{sec:conclusion}

The \CP violation sensitive parameters which describe the \BsDsK decay rates
have been measured using a dataset of $1.0\invfb$ of $pp$ collision data.
Their values are found to be
\begin{align*}
\Cpar  &= \phantom{-}0.53 \pm 0.25 \pm 0.04 \,, \\
\Dpar  &= \phantom{-}0.37 \pm 0.42 \pm 0.20\,, \\
\Dbpar &= \phantom{-}0.20 \pm 0.41 \pm 0.20\,, \\
\Spar  &=           -1.09 \pm 0.33 \pm 0.08\,, \\
\Sbpar &=           -0.36 \pm 0.34 \pm 0.08\,, 
\end{align*}
where the first uncertainties are statistical and the second are systematic.
The results are interpreted in terms of the CKM angle \g, which yields
$\g = (115_{-43}^{+28})^\circ$, $\delta=(3^{+19}_{-20})^{\circ}$ and $r_{\DsK}=0.53^{+0.17}_{-0.16}$ (all angles are given ~modulo~$180^\circ$) at the 68\% confidence level.
This is the first measurement of \g performed in this channel.

% Do not include this in analysis note and conference reports
\section*{Acknowledgements}

\noindent We express our gratitude to our colleagues in the CERN
accelerator departments for the excellent performance of the LHC. We
thank the technical and administrative staff at the LHCb
institutes. We acknowledge support from CERN and from the national
agencies: CAPES, CNPq, FAPERJ and FINEP (Brazil); NSFC (China);
CNRS/IN2P3 (France); BMBF, DFG, HGF and MPG (Germany); SFI (Ireland); INFN (Italy); 
FOM and NWO (The Netherlands); MNiSW and NCN (Poland); MEN/IFA (Romania); 
MinES and FANO (Russia); MinECo (Spain); SNSF and SER (Switzerland); 
NASU (Ukraine); STFC (United Kingdom); NSF (USA).
The Tier1 computing centres are supported by IN2P3 (France), KIT and BMBF 
(Germany), INFN (Italy), NWO and SURF (The Netherlands), PIC (Spain), GridPP 
(United Kingdom).
We are indebted to the communities behind the multiple open 
source software packages on which we depend. We are also thankful for the 
computing resources and the access to software R\&D tools provided by Yandex LLC (Russia).
Individual groups or members have received support from 
EPLANET, Marie Sk\l{}odowska-Curie Actions and ERC (European Union), 
Conseil g\'{e}n\'{e}ral de Haute-Savoie, Labex ENIGMASS and OCEVU, 
R\'{e}gion Auvergne (France), RFBR (Russia), XuntaGal and GENCAT (Spain), Royal Society and Royal
Commission for the Exhibition of 1851 (United Kingdom).

\newpage
\addcontentsline{toc}{section}{References}
\setboolean{inbibliography}{true}
\bibliographystyle{LHCb}
\bibliography{main,LHCb-PAPER,LHCb-CONF,LHCb-DP,LHCb-TDR,ourbib}

\ifx\mcitethebibliography\mciteundefinedmacro
\PackageError{LHCb.bst}{mciteplus.sty has not been loaded}
{This bibstyle requires the use of the mciteplus package.}\fi
\providecommand{\href}[2]{#2}
\begin{mcitethebibliography}{10}
\mciteSetBstSublistMode{n}
\mciteSetBstMaxWidthForm{subitem}{\alph{mcitesubitemcount})}
\mciteSetBstSublistLabelBeginEnd{\mcitemaxwidthsubitemform\space}
{\relax}{\relax}

\bibitem{CKM1}
N.~Cabibbo, \ifthenelse{\boolean{articletitles}}{{\it Unitary symmetry and
  leptonic decays},
  }{}\href{http://dx.doi.org/10.1103/PhysRevLett.10.531}{Phys.\ Rev.\ Lett.\
  {\bf 10} (1963) 531}\relax
\mciteBstWouldAddEndPuncttrue
\mciteSetBstMidEndSepPunct{\mcitedefaultmidpunct}
{\mcitedefaultendpunct}{\mcitedefaultseppunct}\relax
\EndOfBibitem
\bibitem{CKM2}
M.~Kobayashi and T.~Maskawa, \ifthenelse{\boolean{articletitles}}{{\it {CP
  Violation in the Renormalizable Theory of Weak Interaction}},
  }{}\href{http://dx.doi.org/10.1143/PTP.49.652}{Prog.\ Theor.\ Phys.\  {\bf
  49} (1973) 652}\relax
\mciteBstWouldAddEndPuncttrue
\mciteSetBstMidEndSepPunct{\mcitedefaultmidpunct}
{\mcitedefaultendpunct}{\mcitedefaultseppunct}\relax
\EndOfBibitem
\bibitem{Dunietz:1987bv}
I.~Dunietz and R.~G. Sachs, \ifthenelse{\boolean{articletitles}}{{\it
  {Asymmetry Between Inclusive Charmed and Anticharmed Modes in $B^0$,
  Anti-$B^0$ Decay as a Measure of {CP} Violation}},
  }{}\href{http://dx.doi.org/10.1103/PhysRevD.37.3186,
  10.1103/PhysRevD.39.3515}{Phys.\ Rev.\  {\bf D37} (1988) 3186}\relax
\mciteBstWouldAddEndPuncttrue
\mciteSetBstMidEndSepPunct{\mcitedefaultmidpunct}
{\mcitedefaultendpunct}{\mcitedefaultseppunct}\relax
\EndOfBibitem
\bibitem{Aleksan:1991nh}
R.~Aleksan, I.~Dunietz, and B.~Kayser,
  \ifthenelse{\boolean{articletitles}}{{\it {Determining the CP violating phase
  $\gamma$}}, }{}\href{http://dx.doi.org/10.1007/BF01559494}{Z.\ Phys.\  {\bf
  C54} (1992) 653}\relax
\mciteBstWouldAddEndPuncttrue
\mciteSetBstMidEndSepPunct{\mcitedefaultmidpunct}
{\mcitedefaultendpunct}{\mcitedefaultseppunct}\relax
\EndOfBibitem
\bibitem{Fleischer:2003yb}
R.~Fleischer, \ifthenelse{\boolean{articletitles}}{{\it {New strategies to
  obtain insights into CP violation through $B_{(s)} \to D_{(s)}^\pm K^\mp$,
  $D_{(s)}^{*\pm} K^\mp$, ... and $B_{(d)} \to D^\pm \pi^\mp$, $D^{*\pm}
  \pi^\mp$, ... decays}},
  }{}\href{http://dx.doi.org/10.1016/j.nuclphysb.2003.08.010}{Nucl.\ Phys.\
  {\bf B671} (2003) 459}, \href{http://arxiv.org/abs/hep-ph/0304027}{{\tt
  arXiv:hep-ph/0304027}}\relax
\mciteBstWouldAddEndPuncttrue
\mciteSetBstMidEndSepPunct{\mcitedefaultmidpunct}
{\mcitedefaultendpunct}{\mcitedefaultseppunct}\relax
\EndOfBibitem
\bibitem{Aubert:2005yf}
{BaBar collaboration}, B.~Aubert {\em et~al.},
  \ifthenelse{\boolean{articletitles}}{{\it {Measurement of time-dependent
  CP-violating asymmetries and constraints on $\sin(2\beta+\gamma)$ with
  partial reconstruction of $B \to D^{*\mp} \pi^\pm$ decays}},
  }{}\href{http://dx.doi.org/10.1103/PhysRevD.71.112003}{Phys.\ Rev.\  {\bf
  D71} (2005) 112003}, \href{http://arxiv.org/abs/hep-ex/0504035}{{\tt
  arXiv:hep-ex/0504035}}\relax
\mciteBstWouldAddEndPuncttrue
\mciteSetBstMidEndSepPunct{\mcitedefaultmidpunct}
{\mcitedefaultendpunct}{\mcitedefaultseppunct}\relax
\EndOfBibitem
\bibitem{Aubert:2006tw}
{BaBar collaboration}, B.~Aubert {\em et~al.},
  \ifthenelse{\boolean{articletitles}}{{\it {Measurement of time-dependent CP
  asymmetries in $B^0 \to D^{(*)\pm} \pi^\mp$ and $B^0 \to D^\pm \rho^\mp$
  decays}}, }{}\href{http://dx.doi.org/10.1103/PhysRevD.73.111101}{Phys.\ Rev.\
  D {\bf 73} (2006) 111101}, \href{http://arxiv.org/abs/hep-ex/0602049}{{\tt
  arXiv:hep-ex/0602049}}\relax
\mciteBstWouldAddEndPuncttrue
\mciteSetBstMidEndSepPunct{\mcitedefaultmidpunct}
{\mcitedefaultendpunct}{\mcitedefaultseppunct}\relax
\EndOfBibitem
\bibitem{PhysRevD.73.092003}
Belle collaboration, F.~J. Ronga {\em et~al.},
  \ifthenelse{\boolean{articletitles}}{{\it {Measurement of $CP$ violation in
  ${B}^{0} \rightarrow D^{*-} \pi^{+}$ and ${B}^{0} \rightarrow D^{-} \pi^{+}$
  decays}}, }{}\href{http://dx.doi.org/10.1103/PhysRevD.73.092003}{Phys.\ Rev.\
   {\bf D73} (2006) 092003}\relax
\mciteBstWouldAddEndPuncttrue
\mciteSetBstMidEndSepPunct{\mcitedefaultmidpunct}
{\mcitedefaultendpunct}{\mcitedefaultseppunct}\relax
\EndOfBibitem
\bibitem{Bahinipati:2011yq}
Belle collaboration, S.~Bahinipati {\em et~al.},
  \ifthenelse{\boolean{articletitles}}{{\it {Measurements of time-dependent CP
  asymmetries in $B \to D^{*\mp} \pi^{\pm}$ decays using a partial
  reconstruction technique}},
  }{}\href{http://dx.doi.org/10.1103/PhysRevD.84.021101}{Phys.\ Rev.\  {\bf
  D84} (2011) 021101}, \href{http://arxiv.org/abs/1102.0888}{{\tt
  arXiv:1102.0888}}\relax
\mciteBstWouldAddEndPuncttrue
\mciteSetBstMidEndSepPunct{\mcitedefaultmidpunct}
{\mcitedefaultendpunct}{\mcitedefaultseppunct}\relax
\EndOfBibitem
\bibitem{Baak:2007gp}
M.~A. Baak, {\em {Measurement of CKM angle gamma with charmed $B^0$ meson
  decays}}, PhD thesis, Vrije Universiteit Amsterdam, {2007},
  \href{http://www.slac.stanford.edu/pubs/slacreports/slac-r-858.html}{SLAC-R-858}\relax
\mciteBstWouldAddEndPuncttrue
\mciteSetBstMidEndSepPunct{\mcitedefaultmidpunct}
{\mcitedefaultendpunct}{\mcitedefaultseppunct}\relax
\EndOfBibitem
\bibitem{Wolfenstein:1983yz}
L.~Wolfenstein, \ifthenelse{\boolean{articletitles}}{{\it {Parametrization of
  the Kobayashi-Maskawa matrix}},
  }{}\href{http://dx.doi.org/10.1103/PhysRevLett.51.1945}{Phys.\ Rev.\ Lett.\
  {\bf 51} (1983) 1945}\relax
\mciteBstWouldAddEndPuncttrue
\mciteSetBstMidEndSepPunct{\mcitedefaultmidpunct}
{\mcitedefaultendpunct}{\mcitedefaultseppunct}\relax
\EndOfBibitem
\bibitem{PDG}
Particle Data Group, J.~Beringer {\em et~al.},
  \ifthenelse{\boolean{articletitles}}{{\it {Review of Particle Physics}},
  }{}\href{http://dx.doi.org/10.1103/PhysRevD.86.010001}{Phys.\ Rev.\  {\bf
  D86} (2012) 010001}\relax
\mciteBstWouldAddEndPuncttrue
\mciteSetBstMidEndSepPunct{\mcitedefaultmidpunct}
{\mcitedefaultendpunct}{\mcitedefaultseppunct}\relax
\EndOfBibitem
\bibitem{LHCb-PAPER-2013-002}
LHCb collaboration, R.~Aaij {\em et~al.},
  \ifthenelse{\boolean{articletitles}}{{\it {Measurement of $CP$ violation and
  the $B^0_s$ meson decay width difference with $B_s^0\to J/\psi K^+K^-$ and
  $B_s^0 \to J/\psi\pi^+\pi^-$ decays}},
  }{}\href{http://dx.doi.org/10.1103/PhysRevD.87.112010}{Phys.\ Rev.\  {\bf
  D87} (2013) 112010}, \href{http://arxiv.org/abs/1304.2600}{{\tt
  arXiv:1304.2600}}\relax
\mciteBstWouldAddEndPuncttrue
\mciteSetBstMidEndSepPunct{\mcitedefaultmidpunct}
{\mcitedefaultendpunct}{\mcitedefaultseppunct}\relax
\EndOfBibitem
\bibitem{LHCb-PAPER-2011-028}
LHCb collaboration, R.~Aaij {\em et~al.},
  \ifthenelse{\boolean{articletitles}}{{\it {Determination of the sign of the
  decay width difference in the $B^0_s$ system}},
  }{}\href{http://dx.doi.org/10.1103/PhysRevLett.108.241801}{Phys.\ Rev.\
  Lett.\  {\bf 108} (2012) 241801}, \href{http://arxiv.org/abs/1202.4717}{{\tt
  arXiv:1202.4717}}\relax
\mciteBstWouldAddEndPuncttrue
\mciteSetBstMidEndSepPunct{\mcitedefaultmidpunct}
{\mcitedefaultendpunct}{\mcitedefaultseppunct}\relax
\EndOfBibitem
\bibitem{Pivk:2004ty}
M.~Pivk and F.~R. Le~Diberder, \ifthenelse{\boolean{articletitles}}{{\it
  {sPlot: a statistical tool to unfold data distributions}},
  }{}\href{http://dx.doi.org/10.1016/j.nima.2005.08.106}{Nucl.\ Instrum.\
  Meth.\  {\bf A555} (2005) 356},
  \href{http://arxiv.org/abs/physics/0402083}{{\tt
  arXiv:physics/0402083}}\relax
\mciteBstWouldAddEndPuncttrue
\mciteSetBstMidEndSepPunct{\mcitedefaultmidpunct}
{\mcitedefaultendpunct}{\mcitedefaultseppunct}\relax
\EndOfBibitem
\bibitem{2009arXiv0905.0724X}
Y.~{Xie}, \ifthenelse{\boolean{articletitles}}{{\it {sFit: a method for
  background subtraction in maximum likelihood fit}},
  }{}\href{http://arxiv.org/abs/0905.0724}{{\tt arXiv:0905.0724}}\relax
\mciteBstWouldAddEndPuncttrue
\mciteSetBstMidEndSepPunct{\mcitedefaultmidpunct}
{\mcitedefaultendpunct}{\mcitedefaultseppunct}\relax
\EndOfBibitem
\bibitem{Alves:2008zz}
LHCb collaboration, A.~A. Alves~Jr. {\em et~al.},
  \ifthenelse{\boolean{articletitles}}{{\it {The \lhcb detector at the LHC}},
  }{}\href{http://dx.doi.org/10.1088/1748-0221/3/08/S08005}{JINST {\bf 3}
  (2008) S08005}\relax
\mciteBstWouldAddEndPuncttrue
\mciteSetBstMidEndSepPunct{\mcitedefaultmidpunct}
{\mcitedefaultendpunct}{\mcitedefaultseppunct}\relax
\EndOfBibitem
\bibitem{LHCb-DP-2014-001}
R.~Aaij {\em et~al.}, \ifthenelse{\boolean{articletitles}}{{\it {Performance of
  the LHCb Vertex Locator}}, }{}\href{http://arxiv.org/abs/1405.7808}{{\tt
  arXiv:1405.7808}}, {submitted to JINST}\relax
\mciteBstWouldAddEndPuncttrue
\mciteSetBstMidEndSepPunct{\mcitedefaultmidpunct}
{\mcitedefaultendpunct}{\mcitedefaultseppunct}\relax
\EndOfBibitem
\bibitem{LHCb-DP-2013-003}
R.~Arink {\em et~al.}, \ifthenelse{\boolean{articletitles}}{{\it {Performance
  of the LHCb Outer Tracker}},
  }{}\href{http://dx.doi.org/10.1088/1748-0221/9/01/P01002}{JINST {\bf 9}
  (2014) P01002}, \href{http://arxiv.org/abs/1311.3893}{{\tt
  arXiv:1311.3893}}\relax
\mciteBstWouldAddEndPuncttrue
\mciteSetBstMidEndSepPunct{\mcitedefaultmidpunct}
{\mcitedefaultendpunct}{\mcitedefaultseppunct}\relax
\EndOfBibitem
\bibitem{LHCb-DP-2012-003}
M.~Adinolfi {\em et~al.}, \ifthenelse{\boolean{articletitles}}{{\it
  {Performance of the \lhcb RICH detector at the LHC}},
  }{}\href{http://dx.doi.org/10.1140/epjc/s10052-013-2431-9}{Eur.\ Phys.\ J.\
  {\bf C73} (2013) 2431}, \href{http://arxiv.org/abs/1211.6759}{{\tt
  arXiv:1211.6759}}\relax
\mciteBstWouldAddEndPuncttrue
\mciteSetBstMidEndSepPunct{\mcitedefaultmidpunct}
{\mcitedefaultendpunct}{\mcitedefaultseppunct}\relax
\EndOfBibitem
\bibitem{LHCb-DP-2012-004}
R.~Aaij {\em et~al.}, \ifthenelse{\boolean{articletitles}}{{\it {The \lhcb
  trigger and its performance in 2011}},
  }{}\href{http://dx.doi.org/10.1088/1748-0221/8/04/P04022}{JINST {\bf 8}
  (2013) P04022}, \href{http://arxiv.org/abs/1211.3055}{{\tt
  arXiv:1211.3055}}\relax
\mciteBstWouldAddEndPuncttrue
\mciteSetBstMidEndSepPunct{\mcitedefaultmidpunct}
{\mcitedefaultendpunct}{\mcitedefaultseppunct}\relax
\EndOfBibitem
\bibitem{BBDT}
V.~V. Gligorov and M.~Williams, \ifthenelse{\boolean{articletitles}}{{\it
  {Efficient, reliable and fast high-level triggering using a bonsai boosted
  decision tree}},
  }{}\href{http://dx.doi.org/10.1088/1748-0221/8/02/P02013}{JINST {\bf 8}
  (2013) P02013}, \href{http://arxiv.org/abs/1210.6861}{{\tt
  arXiv:1210.6861}}\relax
\mciteBstWouldAddEndPuncttrue
\mciteSetBstMidEndSepPunct{\mcitedefaultmidpunct}
{\mcitedefaultendpunct}{\mcitedefaultseppunct}\relax
\EndOfBibitem
\bibitem{Sjostrand:2006za}
T.~Sj\"{o}strand, S.~Mrenna, and P.~Skands,
  \ifthenelse{\boolean{articletitles}}{{\it {PYTHIA 6.4 physics and manual}},
  }{}\href{http://dx.doi.org/10.1088/1126-6708/2006/05/026}{JHEP {\bf 05}
  (2006) 026}, \href{http://arxiv.org/abs/hep-ph/0603175}{{\tt
  arXiv:hep-ph/0603175}}\relax
\mciteBstWouldAddEndPuncttrue
\mciteSetBstMidEndSepPunct{\mcitedefaultmidpunct}
{\mcitedefaultendpunct}{\mcitedefaultseppunct}\relax
\EndOfBibitem
\bibitem{LHCb-PROC-2010-056}
I.~Belyaev {\em et~al.}, \ifthenelse{\boolean{articletitles}}{{\it {Handling of
  the generation of primary events in \gauss, the \lhcb simulation framework}},
  }{}\href{http://dx.doi.org/10.1109/NSSMIC.2010.5873949}{Nuclear Science
  Symposium Conference Record (NSS/MIC) {\bf IEEE} (2010) 1155}\relax
\mciteBstWouldAddEndPuncttrue
\mciteSetBstMidEndSepPunct{\mcitedefaultmidpunct}
{\mcitedefaultendpunct}{\mcitedefaultseppunct}\relax
\EndOfBibitem
\bibitem{Lange:2001uf}
D.~J. Lange, \ifthenelse{\boolean{articletitles}}{{\it {The EvtGen particle
  decay simulation package}},
  }{}\href{http://dx.doi.org/10.1016/S0168-9002(01)00089-4}{Nucl.\ Instrum.\
  Meth.\  {\bf A462} (2001) 152}\relax
\mciteBstWouldAddEndPuncttrue
\mciteSetBstMidEndSepPunct{\mcitedefaultmidpunct}
{\mcitedefaultendpunct}{\mcitedefaultseppunct}\relax
\EndOfBibitem
\bibitem{Golonka:2005pn}
P.~Golonka and Z.~Was, \ifthenelse{\boolean{articletitles}}{{\it {PHOTOS Monte
  Carlo: a precision tool for QED corrections in $Z$ and $W$ decays}},
  }{}\href{http://dx.doi.org/10.1140/epjc/s2005-02396-4}{Eur.\ Phys.\ J.\  {\bf
  C45} (2006) 97}, \href{http://arxiv.org/abs/hep-ph/0506026}{{\tt
  arXiv:hep-ph/0506026}}\relax
\mciteBstWouldAddEndPuncttrue
\mciteSetBstMidEndSepPunct{\mcitedefaultmidpunct}
{\mcitedefaultendpunct}{\mcitedefaultseppunct}\relax
\EndOfBibitem
\bibitem{Allison:2006ve}
Geant4 collaboration, J.~Allison {\em et~al.},
  \ifthenelse{\boolean{articletitles}}{{\it {Geant4 developments and
  applications}}, }{}\href{http://dx.doi.org/10.1109/TNS.2006.869826}{IEEE
  Trans.\ Nucl.\ Sci.\  {\bf 53} (2006) 270}\relax
\mciteBstWouldAddEndPuncttrue
\mciteSetBstMidEndSepPunct{\mcitedefaultmidpunct}
{\mcitedefaultendpunct}{\mcitedefaultseppunct}\relax
\EndOfBibitem
\bibitem{Agostinelli:2002hh}
Geant4 collaboration, S.~Agostinelli {\em et~al.},
  \ifthenelse{\boolean{articletitles}}{{\it {Geant4: a simulation toolkit}},
  }{}\href{http://dx.doi.org/10.1016/S0168-9002(03)01368-8}{Nucl.\ Instrum.\
  Meth.\  {\bf A506} (2003) 250}\relax
\mciteBstWouldAddEndPuncttrue
\mciteSetBstMidEndSepPunct{\mcitedefaultmidpunct}
{\mcitedefaultendpunct}{\mcitedefaultseppunct}\relax
\EndOfBibitem
\bibitem{LHCb-PROC-2011-006}
M.~Clemencic {\em et~al.}, \ifthenelse{\boolean{articletitles}}{{\it {The \lhcb
  simulation application, \gauss: design, evolution and experience}},
  }{}\href{http://dx.doi.org/10.1088/1742-6596/331/3/032023}{{J.\ Phys.\ Conf.\
  Ser.\ } {\bf 331} (2011) 032023}\relax
\mciteBstWouldAddEndPuncttrue
\mciteSetBstMidEndSepPunct{\mcitedefaultmidpunct}
{\mcitedefaultendpunct}{\mcitedefaultseppunct}\relax
\EndOfBibitem
\bibitem{Hulsbergen2005566}
W.~D. Hulsbergen, \ifthenelse{\boolean{articletitles}}{{\it {Decay chain
  fitting with a Kalman filter}},
  }{}\href{http://dx.doi.org/http://dx.doi.org/10.1016/j.nima.2005.06.078}{NIMA
  {\bf 552} (2005), no.~3 566 }\relax
\mciteBstWouldAddEndPuncttrue
\mciteSetBstMidEndSepPunct{\mcitedefaultmidpunct}
{\mcitedefaultendpunct}{\mcitedefaultseppunct}\relax
\EndOfBibitem
\bibitem{Breiman}
L.~Breiman, J.~H. Friedman, R.~A. Olshen, and C.~J. Stone, {\em Classification
  and regression trees}, Wadsworth international group, Belmont, California,
  USA, 1984\relax
\mciteBstWouldAddEndPuncttrue
\mciteSetBstMidEndSepPunct{\mcitedefaultmidpunct}
{\mcitedefaultendpunct}{\mcitedefaultseppunct}\relax
\EndOfBibitem
\bibitem{TMVA}
A.~Hocker {\em et~al.}, \ifthenelse{\boolean{articletitles}}{{\it {TMVA -
  toolkit for multivariate data analysis}}, }{}PoS {\bf ACAT} (2007) 040,
  \href{http://arxiv.org/abs/physics/0703039}{{\tt
  arXiv:physics/0703039}}\relax
\mciteBstWouldAddEndPuncttrue
\mciteSetBstMidEndSepPunct{\mcitedefaultmidpunct}
{\mcitedefaultendpunct}{\mcitedefaultseppunct}\relax
\EndOfBibitem
\bibitem{LHCb-PROC-2011-008}
A.~Powell {\em et~al.}, \ifthenelse{\boolean{articletitles}}{{\it {Particle
  identification at LHCb}}, }{}PoS {\bf ICHEP2010} (2010) 020,
  \href{https://cdsweb.cern.ch/record/1322666?ln=en}{LHCb-PROC-2011-008}\relax
\mciteBstWouldAddEndPuncttrue
\mciteSetBstMidEndSepPunct{\mcitedefaultmidpunct}
{\mcitedefaultendpunct}{\mcitedefaultseppunct}\relax
\EndOfBibitem
\bibitem{Skwarnicki:1986xj}
T.~Skwarnicki, {\em {A study of the radiative cascade transitions between the
  Upsilon-prime and Upsilon resonances}}, PhD thesis, Institute of Nuclear
  Physics, Krakow, 1986,
  {\href{http://inspirehep.net/record/230779/files/230779.pdf}{DESY-F31-86-02}}\relax
\mciteBstWouldAddEndPuncttrue
\mciteSetBstMidEndSepPunct{\mcitedefaultmidpunct}
{\mcitedefaultendpunct}{\mcitedefaultseppunct}\relax
\EndOfBibitem
\bibitem{Cranmer:2000du}
K.~S. Cranmer, \ifthenelse{\boolean{articletitles}}{{\it {Kernel estimation in
  high-energy physics}},
  }{}\href{http://dx.doi.org/10.1016/S0010-4655(00)00243-5}{Comput.\ Phys.\
  Commun.\  {\bf 136} (2001) 198},
  \href{http://arxiv.org/abs/hep-ex/0011057}{{\tt arXiv:hep-ex/0011057}}\relax
\mciteBstWouldAddEndPuncttrue
\mciteSetBstMidEndSepPunct{\mcitedefaultmidpunct}
{\mcitedefaultendpunct}{\mcitedefaultseppunct}\relax
\EndOfBibitem
\bibitem{Krocker}
G.~A. Krocker, {\em {Development and calibration of a same side kaon tagging
  algorithm and measurement of the $B^{0}_{s}$-$\bar{B}^{0}_{s}$ oscillation
  frequency $\Delta m_{s}$ at the LHCb experiment}}, PhD thesis, Heidelberg U.,
  Sep, 2013,
  \href{https://cds.cern.ch/record/1631104?ln=en}{CERN-THESIS-2013-213}\relax
\mciteBstWouldAddEndPuncttrue
\mciteSetBstMidEndSepPunct{\mcitedefaultmidpunct}
{\mcitedefaultendpunct}{\mcitedefaultseppunct}\relax
\EndOfBibitem
\bibitem{LHCb-CONF-2012-033}
{LHCb collaboration}, \ifthenelse{\boolean{articletitles}}{{\it {Optimization
  and calibration of the same-side kaon tagging algorithm using hadronic $\Bs$
  decays in 2011 data}}, }{}
  \href{http://cdsweb.cern.ch/search?p={LHCb-CONF-2012-033}&f=reportnumber&action_search=Search&c=LHCb+Reports&c=LHCb+Conference+Proceedings&c=LHCb+Conference+Contributions&c=LHCb+Notes&c=LHCb+Theses&c=LHCb+Papers}
  {{LHCb-CONF-2012-033}}\relax
\mciteBstWouldAddEndPuncttrue
\mciteSetBstMidEndSepPunct{\mcitedefaultmidpunct}
{\mcitedefaultendpunct}{\mcitedefaultseppunct}\relax
\EndOfBibitem
\bibitem{LHCb-PAPER-2013-006}
LHCb collaboration, R.~Aaij {\em et~al.},
  \ifthenelse{\boolean{articletitles}}{{\it {Precision measurement of the
  $B^0_s-\bar{B}^0_s$ oscillation frequency in the decay $B^0_s \to D^+_s
  \pi^-$}}, }{}\href{http://dx.doi.org/10.1088/1367-2630/15/5/053021}{New J.\
  Phys.\  {\bf 15} (2013) 053021}, \href{http://arxiv.org/abs/1304.4741}{{\tt
  arXiv:1304.4741}}\relax
\mciteBstWouldAddEndPuncttrue
\mciteSetBstMidEndSepPunct{\mcitedefaultmidpunct}
{\mcitedefaultendpunct}{\mcitedefaultseppunct}\relax
\EndOfBibitem
\bibitem{Karbach:ComplxErrFunc}
M.~Karbach, G.~Raven, and M.~Schiller,
  \ifthenelse{\boolean{articletitles}}{{\it Decay time integrals in neutral
  meson mixing and their efficient evaluation},
  }{}\href{http://arxiv.org/abs/1407.0748}{{\tt arXiv:1407.0748}}\relax
\mciteBstWouldAddEndPuncttrue
\mciteSetBstMidEndSepPunct{\mcitedefaultmidpunct}
{\mcitedefaultendpunct}{\mcitedefaultseppunct}\relax
\EndOfBibitem
\bibitem{LHCb-PAPER-2014-003}
LHCb collaboration, R.~Aaij {\em et~al.},
  \ifthenelse{\boolean{articletitles}}{{\it {Precision measurement of the ratio
  of the $\Lambda_b^0$ to $\bar{B}^0$ lifetimes}},
  }{}\href{http://dx.doi.org/10.1016/j.physletb.2014.05.021}{Phys.\ Lett.\
  {\bf B734} (2014) 122}, \href{http://arxiv.org/abs/1402.6242}{{\tt
  arXiv:1402.6242}}\relax
\mciteBstWouldAddEndPuncttrue
\mciteSetBstMidEndSepPunct{\mcitedefaultmidpunct}
{\mcitedefaultendpunct}{\mcitedefaultseppunct}\relax
\EndOfBibitem
\bibitem{LHCB-PAPER-2014-013}
LHCb collaboration, R.~Aaij {\em et~al.},
  \ifthenelse{\boolean{articletitles}}{{\it {Measurement of $CP$ asymmetry in
  $D^0 \to K^- K^+$ and $D^0 \to \pi^- \pi^+$ decays}},
  }{}\href{http://dx.doi.org/10.1007/JHEP07(2014)041}{JHEP {\bf 07} (2014)
  041}, \href{http://arxiv.org/abs/1405.2797}{{\tt arXiv:1405.2797}}\relax
\mciteBstWouldAddEndPuncttrue
\mciteSetBstMidEndSepPunct{\mcitedefaultmidpunct}
{\mcitedefaultendpunct}{\mcitedefaultseppunct}\relax
\EndOfBibitem
\bibitem{LHCB-PAPER-2013-053}
LHCb collaboration, R.~Aaij {\em et~al.},
  \ifthenelse{\boolean{articletitles}}{{\it {Measurement of $D^0$--$\bar{D}^0$
  mixing parameters and search for $CP$ violation using $D^0\to K^+\pi^-$
  decays}}, }{}\href{http://dx.doi.org/10.1103/PhysRevLett.111.251801}{Phys.\
  Rev.\ Lett.\  {\bf 111} (2013) 251801},
  \href{http://arxiv.org/abs/1309.6534}{{\tt arXiv:1309.6534}}\relax
\mciteBstWouldAddEndPuncttrue
\mciteSetBstMidEndSepPunct{\mcitedefaultmidpunct}
{\mcitedefaultendpunct}{\mcitedefaultseppunct}\relax
\EndOfBibitem
\bibitem{LHCb-PAPER-2013-020}
LHCb collaboration, R.~Aaij {\em et~al.},
  \ifthenelse{\boolean{articletitles}}{{\it {A measurement of the CKM angle
  $\gamma$ from a combination of $B^\pm \to Dh^\pm$ analyses}},
  }{}\href{http://dx.doi.org/10.1016/j.physletb.2013.08.020}{Phys.\ Lett.\
  {\bf B726} (2013) 151}, \href{http://arxiv.org/abs/1305.2050}{{\tt
  arXiv:1305.2050}}\relax
\mciteBstWouldAddEndPuncttrue
\mciteSetBstMidEndSepPunct{\mcitedefaultmidpunct}
{\mcitedefaultendpunct}{\mcitedefaultseppunct}\relax
\EndOfBibitem
\end{mcitethebibliography}
\newpage
%%%%%%%%%%%%%%%%%%%%%%%%%%%%%%%%%%%%%%%%%%
\centerline{\large\bf LHCb collaboration}
\begin{flushleft}
\small
R.~Aaij$^{41}$, 
B.~Adeva$^{37}$, 
M.~Adinolfi$^{46}$, 
A.~Affolder$^{52}$, 
Z.~Ajaltouni$^{5}$, 
S.~Akar$^{6}$, 
J.~Albrecht$^{9}$, 
F.~Alessio$^{38}$, 
M.~Alexander$^{51}$, 
S.~Ali$^{41}$
% ...

{\footnotesize \it
% ...
$ ^{5}$Clermont Universit\'{e}, Universit\'{e} Blaise Pascal, CNRS/IN2P3, LPC, Clermont-Ferrand, France\\
$ ^{6}$CPPM, Aix-Marseille Universit\'{e}, CNRS/IN2P3, Marseille, France\\
% ...
$ ^{9}$Fakult\"{a}t Physik, Technische Universit\"{a}t Dortmund, Dortmund, Germany\\
% ...
$ ^{37}$Universidad de Santiago de Compostela, Santiago de Compostela, Spain\\
$ ^{38}$European Organization for Nuclear Research (CERN), Geneva, Switzerland\\
% ...
$ ^{41}$Nikhef National Institute for Subatomic Physics, Amsterdam, The Netherlands\\
% ...
$ ^{46}$H.H. Wills Physics Laboratory, University of Bristol, Bristol, United Kingdom\\
% ...
$ ^{51}$School of Physics and Astronomy, University of Glasgow, Glasgow, United Kingdom\\
$ ^{52}$Oliver Lodge Laboratory, University of Liverpool, Liverpool, United Kingdom\\
% ...
}
\end{flushleft}
%%%%%%%%%%%%%%%%%%%%%%%%%%%%%%%%%%%%%%%%%%

\end{document}